\documentclass[authoryear,reqno,12pt,a4paper]{elsarticle}  
\usepackage{amsthm,amsmath,amssymb,setspace,tabularx,rotating}
\usepackage{amsmath,natbib,graphicx,soul}

\usepackage[export]{adjustbox}
\usepackage{verbatim,booktabs}
\usepackage{algorithm}
\usepackage{algpseudocode}
\usepackage{amsmath}
\usepackage{amsfonts}
\usepackage{amssymb}
\usepackage{svg,orcidlink}
\usepackage{pifont}
\usepackage{ragged2e}
\usepackage{algpseudocode}
\usepackage{caption}
\usepackage{xcolor,hyperref}
\usepackage[toc,page]{appendix}

\usepackage{tabularx}
\usepackage{longtable}
\usepackage[toc,page]{appendix}

\usepackage[top=1in, bottom=1in, left=1in, right=1in]{geometry}
\usepackage{algpseudocode}
\usepackage{lscape}
\usepackage{pdflscape}

\hypersetup{
pdftitle={xxx},
pdfauthor={xxx},
pdfkeywords={xxx},
colorlinks=true,
linkcolor=blue,
citecolor=blue,
urlcolor=blue,
bookmarksnumbered=true,
pdfstartview=
}

\robustify\bfseries
\makeatletter
\let\c@table\c@figure 
\makeatother

\numberwithin{equation}{section}

\newcommand{\citewiki}[1]{(\href{https://en.wikipedia.org/wiki/#1}{\textmd{\textsc{WikipediA}}})}

\journal{Journal of Portfolio Management} %
\begin{document}
\begin{frontmatter}
\renewcommand{\thefootnote}{\fnsymbol{footnote}}
\title{\large Beyond Prompting: Autonomous Factor Investing via Agentic AI
\footnote{Project homepage: \url{https://allenh16.github.io/agentic-factor-investing/}}
\footnote{The authors thank Ivan Blanco (CUNEF), Kaiqi Hu (Rutgers Business School),  Bolong Wang (CITIC Securities), Yifan Ye (BNBU), Chao Zhang (HKUST Guangzhou), Yi Zhang (HKUST Guangzhou), and Yibin Zhang (Bosera Asset Management), and internal seminar participants at X Asset Management for helpful discussions, comments, and support.
These discussions, particularly those drawing upon practical industry insights, were conducted solely for academic purposes; the views expressed are those of the individuals and do not necessarily represent their employers.
The authors also thank the editorial teams of QuantML and LLMQuant  for their insightful coverage and summary of this research across their leading Chinese practitioner-oriented social media platforms, which helped facilitate broader academic and industry exchange.
Any remaining errors or oversights are the responsibility of the authors.}
}
\author[gsm,ust,ustgz]{Allen Yikuan Huang \orcidlink{0000-0002-4881-0972}}
\ead{yhuangfi@connect.ust.hk}
\author[ust,ustgz]{Zheqi Fan \orcidlink{0000-0003-4752-9019}\corref{corrauthor}}
\ead{zheqi.fan@connect.ust.hk}

\address[gsm]{
\Centering Guanghua School of Management,
Peking University, Beijing, China}
\address[ust]{
\Centering Division of EMIA,
Hong Kong University of Science and Technology, Hong Kong SAR}
\address[ustgz]{
\Centering Thrust of FinTech,
Hong Kong University of Science and Technology, Guangzhou, China}
\cortext[corrauthor]{Corresponding author 
}
\date{\today}
\begin{abstract}
\small
{This paper develops an autonomous framework for systematic factor investing via agentic AI.} 
Rather than relying on sequential manual prompts, our approach operationalizes the model as a self-directed engine that endogenously formulates interpretable trading signals. To mitigate data snooping biases, this closed-loop system imposes strict empirical discipline through out-of-sample validation and economic rationale requirements. Applying this methodology to the U.S. equity market, we document that long-short portfolios formed on the simple linear combination of signals deliver an annualized Sharpe ratio of 2.75 and a return of 54.81\%. 
Finally, our empirics demonstrate that self-evolving AI offers a scalable and interpretable paradigm.
\newline


\noindent \textbf{Practical implications}:
\begin{itemize}
    \item Autonomous Alpha Generation: Introduces an Agentic AI framework that acts as a quant researcher to iteratively discover economically interpretable factors.  
    \item Overfitting Mitigation: Systematically combats data mining and p-hacking by enforcing strict out-of-sample testing and requiring clear economic rationale.
    \item Robust Practical Returns: Delivers significant risk-adjusted performance that remains economically meaningful net of realistic transaction costs and turnover constraints.
\end{itemize}
\end{abstract}
\end{frontmatter}
\newpage
\section{Introduction}
\noindent  
The landscape of quantitative investment is undergoing a profound transformation, driven by the rapid evolution of artificial intelligence (AI) and machine learning. Historically, the discovery of systematic return predictors—often referred to as the “factor zoo”—has relied heavily on human intuition, economic hypotheses, and labor-intensive manual testing. While traditional machine learning models have significantly enhanced our ability to process high-dimensional datasets and capture non-linear relationships, they remain fundamentally passive tools. They require researchers to manually engineer features, define explicit rules, and continuously prompt the models to generate insights. This human-dependent paradigm not only creates a severe bottleneck in research efficiency but also structurally exacerbates the risks of data mining and backtest overfitting.

\begin{figure}[htbp]
\begin{center}
\includegraphics[width=0.75\textwidth]{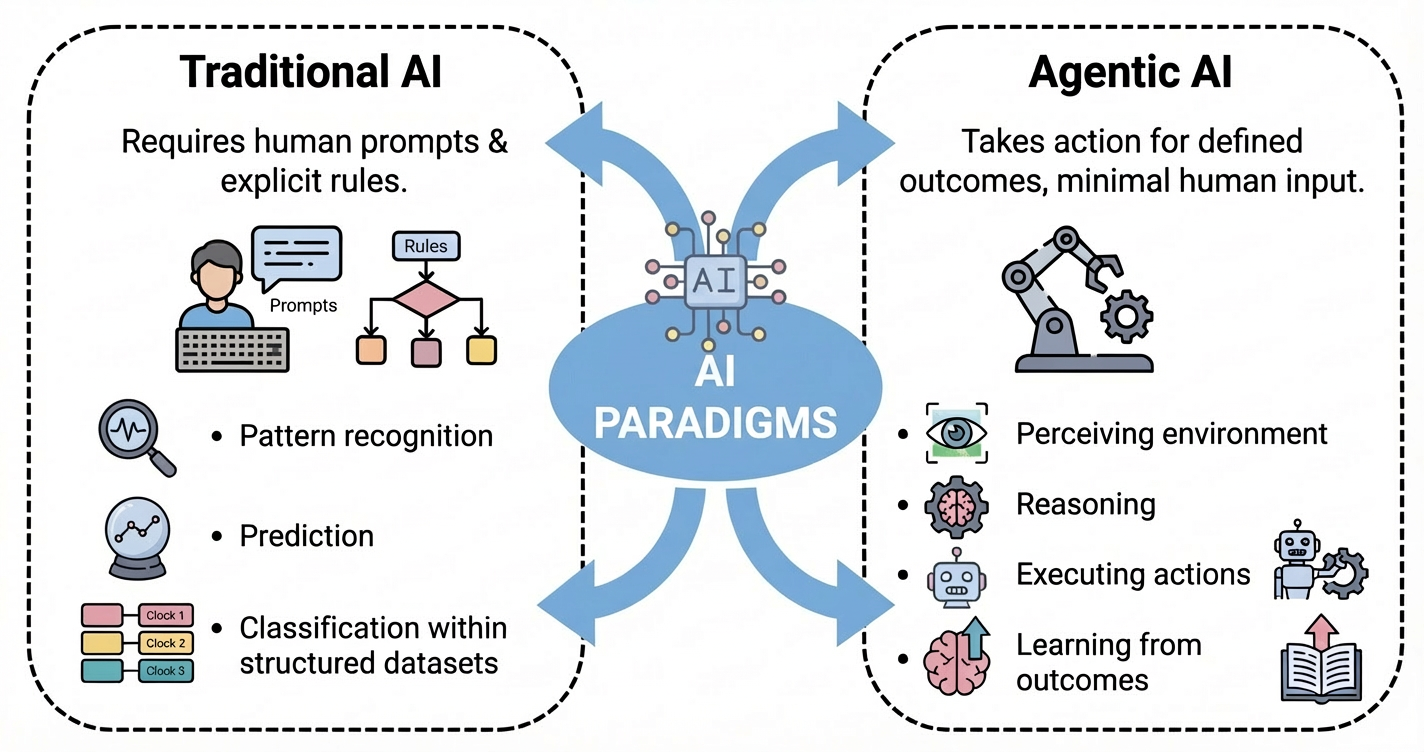}
\caption{\textbf{Core Capabilities Comparison: Traditional vs. Agentic AI}}
\label{fig:agenticAI}
\end{center}
{
\footnotesize
\noindent
\begin{spacing}{1.4}
{This figure illustrates the transition from passive AI systems, which rely on human prompts and explicit rules, to autonomous agentic systems. While traditional AI focuses on pattern recognition and prediction within structured datasets, agentic AI is characterized by its ability to perceive environments, reason, execute actions, and learn from outcomes to achieve defined goals independently.}
\end{spacing}
}
\end{figure}

To overcome these limitations, this paper introduces a novel framework that leverages the paradigm shift from Traditional AI to Agentic AI within the context of quantitative portfolio management. As illustrated in Exhibit \ref{fig:agenticAI}, traditional AI systems function primarily as passive analytical engines. They excel at pattern recognition, prediction, and classification within structured datasets, but their utility is strictly bounded by human prompts and predefined rules. In contrast, Agentic AI represents a leap toward autonomous, goal-driven systems. Rather than merely answering queries, an Agentic AI system perceives its environment (e.g., market dynamics), engages in multi-step reasoning, executes actions, and continuously learns from the outcomes of its decisions with minimal human intervention.

We apply this autonomous agentic framework to the complex domain of alpha generation and factor mining. In our proposed system, the Agentic AI acts as an autonomous quantitative researcher. Given a high-level objective—such as maximizing risk-adjusted returns while strictly controlling for turnover and transaction costs—the agent independently navigates a vast hypothesis space. It formulates novel mathematical expressions for alpha factors using symbolic regression, tests them against historical market data, assesses their economic rationale, and dynamically updates its internal memory based on empirical feedback.

This transition from a ``prompt-driven'' to an ``autonomous-agent'' workflow offers profound advantages for portfolio managers. First, it systematically mitigates human behavioral biases and the pervasive issue of \(p\)-hacking. The agent evaluates factors based on rigorous, out-of-sample statistical properties and structural robustness rather than mere narrative appeal. Second, to translate these machine-generated signals into a cohesive, investable portfolio, our framework integrates advanced non-linear aggregation techniques (e.g., LightGBM). This two-stage approach—autonomous signal discovery followed by non-linear synthesis—ensures that the final portfolio captures complex factor interactions while remaining practically executable.

In this paper, we demonstrate the empirical efficacy of the Agentic AI framework using extensive historical market data. We show that the autonomously generated factor portfolios yield highly significant risk-adjusted returns (alpha) that cannot be explained by traditional asset pricing models. Crucially for practitioners, we rigorously evaluate the strategy's out-of-sample performance net of realistic transaction costs, market microstructure impacts, and turnover constraints. Our findings suggest that Agentic AI is not merely a theoretical computer science concept, but a transformative, scalable engine for modern quantitative asset management—one capable of discovering resilient, investable sources of alpha that traditional human-led research might overlook.

We begin our empirical evaluation by isolating the performance of individual agent-generated signals. Decile portfolio sorts reveal that the top AI-discovered factors generate statistically significant and economically meaningful long-short return spreads. Crucially, these signals exhibit broadly monotonic rank ordering for several factors and deliver positive risk-adjusted returns for the stronger signals against standard asset pricing models (e.g., the Fama-French factors). This confirms that the agent does not merely repackage known risk premia, but autonomously discovers genuine, standalone alpha from raw market data.

Because individual signals often capture distinct, complementary dimensions of market inefficiencies, we next evaluate their efficacy within a cohesive multi-factor portfolio. To aggregate the discovered signals, we evaluate both a transparent linear combination and a nonlinear LightGBM integrator. The linear specification serves as the main benchmark because it is simple, interpretable, and easy to audit, while LightGBM is used as a complementary specification to capture interaction effects and conditional nonlinearities. This design allows us to separate the economic value of the discovered factors from the incremental contribution of model flexibility. The resulting composite strategy delivers encouraging out-of-sample performance. More importantly for institutional practitioners, this outperformance retains substantial performance net of transaction costs, market impact, and strict turnover constraints, proving the strategy's viability for real-world capital deployment.

A pervasive challenge in modern quantitative finance---particularly within automated machine learning---is the acute risk of backtest overfitting and the proliferation of the ``factor zoo.'' To directly address this, we subject our framework to stringent tests for false discoveries. By applying multiple hypothesis testing adjustments and evaluating out-of-sample signal decay, we demonstrate that the Agentic AI systematically resists \(p\)-hacking. The agent's requirement to articulate economic rationale, combined with its dynamic memory updates, enforces a strict research discipline. This ensures that the generated alphas are rooted in persistent structural market dynamics rather than spurious data mining.

Finally, we subject our findings to an extensive battery of robustness checks to confirm their reliability across varying market environments. The Agentic AI-driven portfolios maintain consistent profitability across distinct market regimes, demonstrating resilience during periods of elevated volatility and macroeconomic stress. Furthermore, the performance holds firm across alternative asset universes, extended holding periods, and varied hyperparameter specifications. This comprehensive validation provides institutional investors with confidence that the autonomous alpha generation is structurally sound, regime-resilient, and not an artifact of sample-specific curve fitting.

Ultimately, our core contribution extends beyond merely deploying AI for empirical factor mining. We propose an auditable and economically constrained paradigm for automated investment research. This framework directly addresses two prominent concerns in modern empirical finance: the proliferation of spurious signals in the ``factor zoo'' \citep{cochrane2011presidential, harvey2016cross} and the opacity of black-box machine learning models. By enforcing economic logic at the generation stage and maintaining transparent, human-readable audit trails, our agentic approach bridges the gap between unconstrained computational power and rigorous financial intuition.

The remainder of this paper is organized as follows. Section \ref{sec:literature} reviews the related literature on financial factor models, machine learning in asset pricing, and the application of large language models in finance. Section \ref{sec:Methodology} details the methodology of our Agentic AI framework, outlining the closed-loop workflow and evaluation protocols for autonomous factor generation. Section \ref{sec:data} describes the data, sample construction, and the predictor set. 
Section \ref{sec:empirics} presents the core empirical results, evaluating the performance of both single-factor portfolios and their multivariate combinations. 
Section \ref{sec:mitigation} addresses data-mining concerns by examining the economic rationale, statistical hurdles, and structural anatomy of the agent-generated factors. 
Section \ref{sec:robustness} provides extensive robustness tests, including longer-horizon predictability, transaction costs, turnover analysis, and a comparison against traditional AI frameworks. Finally, Section \ref{sec:Conclusion} concludes the paper.

\section{Related literature}
\label{sec:literature}
\noindent This paper contributes to the vast literature by situating itself at the nexus of three primary research streams: the traditional search for empirical anomalies and technical signals, the application of machine learning in cross-sectional asset pricing, and the emerging frontier of Large Language Models (LLMs) and agentic AI in financial research. By synthesizing these domains, we move beyond static factor discovery toward an autonomous, closed-loop research engine that bridges the gap between human-led hypothesis formulation and automated empirical validation.

\subsection{Financial Factor Models and Technical Signals}
\noindent The quest to explain the cross-section of expected returns began with the Capital Asset Pricing Model (CAPM, \cite{sharpe1964capital}) and evolved into the multi-factor frameworks of \cite{fama1993common, fama2015five} and \cite{carhart1997persistence}. 
While fundamental accounting-based factors dominated early research, a parallel and robust literature explored the predictive power of price-based anomalies and market microstructure signals. \cite{brock1992simple} and \cite{lo2000foundations} provided foundational evidence that technical analysis, such as moving average rules and chart patterns, possesses significant predictive power that cannot be fully explained by random walk hypotheses. Building on this, \cite{han2013new} demonstrated that technical signals are particularly profitable in portfolios characterized by high volatility and high information uncertainty, suggesting that price-volume data captures investor overreaction or underreaction to news. \cite{han2016trend} further unified these insights by constructing a ``trend factor'' that captures short-, intermediate-, and long-term price signals, justifying its existence through a general equilibrium model where investors have heterogeneous beliefs. 
Recent evidence further extends these findings, showing that technical signals serve as a powerful sentiment barometer \citep{ding2023technical}, capture cross-stock predictability \citep{chen2025cross}, predict returns via moving average deviations \citep{ko2025short}, and even outperform traditional characteristics in the corporate bond market \citep{chin2025technical}.
Collectively, these advancements reinforce the growing synergy between academic research and the implementation of systematic, rules-based investment strategies in practice \citep{giamouridis2017systematic}.
Despite their empirical success, these traditional models rely heavily on human-defined rules and static factor constructions. Such manual heuristics are increasingly limited by the ``Factor Zoo'' problem \citep{cochrane2011presidential}, where the sheer dimensionality of potential price-volume combinations exceeds human cognitive capacity and fails to adapt to rapidly evolving market environments.

Relative to this strand of literature, our paper moves beyond manually defined or theoretically-derived static factors. We establish an autonomous agentic loop that discovers, tests, and refines factor hypotheses dynamically. By focusing on raw price and volume data, our framework allows for the emergence of complex, non-linear technical signals that are not pre-constrained by traditional linear factor theory.

\subsection{Machine Learning in Asset Pricing}

\noindent The transition from linear shrinkage models to machine learning (ML) has fundamentally reshaped the landscape of empirical asset pricing\footnote{We refer the readers to \citet{giglio2022factor,kelly2023financial} for comprehensive surveys that synthesize the recent advancements and methodological shifts in this strand of literature.}. 
This evolution mirrors the shift in industry practice toward systematic, data-driven investment frameworks that leverage ML to integrate diverse predictive signals \citep{giamouridis2017systematic}.\footnote{\cite{cerniglia2020selecting} emphasize that this shift does not create a conflict with traditional econometrics, but rather provides a complementary toolkit for integrating financial theory with computational innovation.
For a comprehensive treatment of engineering protocols and practical implementation challenges in financial machine learning that complements the broader literature, see \cite{de2018advances}.}
\cite{gu2020empirical} provided a landmark comparison, showing that deep neural networks and gradient-boosted trees significantly outperform traditional OLS regressions by capturing latent non-linear risk exposures and high-dimensional interactions. Further innovations, such as the Instrumented Principal Component Analysis (IPCA) by \cite{kelly2019characteristics}, allowed latent factors to depend on observable firm characteristics in a time-varying manner, which offers a more dynamic view of risk premia. 

However, the application of ML in asset pricing is not without significant challenges. As noted by \cite{avramov2023machine}, many ML models are highly susceptible to overfitting in the presence of low signal-to-noise ratios and often lack economic interpretability, functioning as black boxes that offer little insight into the underlying economic drivers. Moreover, the iron law of active management, involving transaction costs and capacity constraints, often erodes the gains from high-frequency ML strategies, as documented by \cite{novy2016taxonomy}. A recurring limitation in this literature, echoing the factor zoo problem, is that the information set for feature engineering is still largely pre-selected by humans. While the ML model optimizes the weights, the functional form of the features remains fixed. Early attempts to bypass this human constraint utilized genetic programming to search for optimal trading rules in a non-differentiable functional space \citep{neely1997technical}. Recent evidence from \cite{brogaard2023machine} further suggests that such evolutionary algorithms can outperform strict loss-minimization ML by directly optimizing economic objectives and maintaining flexibility in complex search spaces.\footnote{Alternatively, neural networks can accelerate structural estimation of derivatives pricing models by leveraging their universal approximation properties \citep{ye2025modeling,fan2025options,fan2026deep}.}

Our paper contributes to this field by reframing factor discovery as an automated machine learning (AutoML) and symbolic regression problem. While sharing the goal of exploring functional spaces with genetic programming, our approach leverages the semantic reasoning of large language models to overcome the search efficiency and interpretability limits often associated with traditional evolutionary heuristics. 
This generally aligns with the broader industry trend of integrating AI foundation models into asset management \citep{fabozzi2025implementing}. 
We categorize our approach as a next-generation ML framework where the learner is an LLM-based agent that optimizes over the space of mathematical functional forms. This addresses the black-box concern by producing explicit, interpretable factor formulas while subjecting them to transparent selection gates and explicit anti-overfitting constraints through an iterative empirical feedback loop.

\subsection{Large Language Models and Agentic AI in Finance}
\noindent The integration of Natural Language Processing (NLP) in finance has progressed from simple dictionary-based sentiment analysis \citep{tetlock2007giving, loughran2011liability} to the sophisticated reasoning capabilities of Large Language Models (LLMs). \cite{lopez2023can} demonstrated that ChatGPT can predict stock returns by analyzing news headlines, significantly outperforming traditional sentiment scores. 
Building on this, \cite{chen2022expected} demonstrate that high-dimensional embeddings extracted from financial news by LLMs (such as ChatGPT and LLaMA) capture deep semantic information that substantially improves cross-sectional return predictability across global markets.
This predictive capacity is further refined by \cite{chen2025chatgpt}, who provide a comparative analysis of ChatGPT and DeepSeek, noting that ChatGPT's extensive training allows for more effective extraction of macro-level signals from English financial news.

Beyond textual analysis, recent studies have begun to use LLM architectures to encode structural financial data. \cite{chai2025generative} introduce Risk Premia BERT (RPBERT), which reframes the market cross-section as characteristic-ordered sequences to learn context-dependent firm representations. To ensure the empirical validity of such models, \cite{he2025chronologically} develop ChronoBERT, emphasizing chronological consistency to mitigate lookahead bias and training leakage, which are critical for credible backtesting in financial domains. This creative and inferential capacity was further exploited by \cite{cheng2024gpt}, who showed that GPT-4 can autonomously generate high-return factors through knowledge inference. This methodology was extended to specialized markets, such as the Chinese futures market \citep{cheng2026large}, proving the cross-asset applicability of LLM-generated insights. 
For a comprehensive review of LLM applications in finance, see \cite{kong2024large1,kong2024large2}\footnote{Their survey papers categorize the literature into several key domains, including linguistic tasks, sentiment analysis, financial time series, financial reasoning, and agent-based modeling.}.
Most recently, \cite{pu2026autonomous} introduced a fully agentic nowcasting framework that utilizes real-time web search to evaluate stock attractiveness.
From an institutional perspective, \cite{chen2025agentic} examines how agentic AI can redefine the operating models of asset management through autonomous data analysis and investment decision-making.
However, the transition to fully autonomous AI agents introduces new challenges. 
As documented by \cite{xu2025how}, AI agents can mislead investors through a reliance on outdated training data or web-searching bias, highlighting the reliability risks in automated investment advice.

Relative to these studies, our paper advances from treating AI as a one-off factor generator or nowcaster toward a fully operationalized, closed-loop research engine. Unlike existing frameworks that produce static outputs, our agentic system autonomously proposes construction logic, executes code for backtesting, and adaptively refines its search space based on empirical performance. By automating this iterative cycle of hypothesis formulation, code execution, and empirical validation, we transform the role of AI from a passive assistant into an autonomous researcher. This empirical grounding directly mitigates the aforementioned reliability risks by ensuring that all generated insights are strictly validated against historical market data. We thus create an extensible, logically transparent, and empirically accountable factor library.
To the best of our knowledge, this is the first study, at least in the finance literature, to apply agentic AI to factor investing.

\section{Methodology: Agentic AI for Factor Generation}
\label{sec:Methodology}
\noindent This section describes the methodological framework for systematic factor discovery. The process integrates hypothesis generation with empirical validation through an iterative research cycle. By imposing a fixed set of functional primitives and standardized evaluation protocols, the framework maintains statistical discipline and ensures reproducibility. This structure formalizes the search for asset pricing signals into an auditable sequence where every candidate follows a consistent path from initial specification to final selection.

\subsection{System Design and Closed-Loop Workflow}
\noindent We design the agent as an autonomous research system for systematic factor investing, rather than as a one-shot language interface. The key design goal is to convert factor discovery into a reproducible closed loop with stable rules. In each round, the system generates hypotheses, computes factor values, evaluates candidates under a fixed protocol, applies promotion gates, and updates the next search policy using accumulated evidence. This closed-loop structure is important because it separates autonomous exploration from ad hoc discretion: the agent can search broadly, but it must pass through the same measurement and selection mechanism in every round. 
Formally, we define the discovery cycle at iteration \(k\) as a sequence of functional mappings:
\begin{equation}
\mathcal{C}_k: \mathcal{H}_k \xrightarrow{\text{Gen}} \mathcal{F}_k \xrightarrow{\text{Eval}} \mathcal{M}_k \xrightarrow{\text{Gate}} \mathcal{D}_k
\end{equation}
where \(\mathcal{H}_k\) represents the set of hypotheses, \(\mathcal{F}_k\) the generated factor candidates, \(\mathcal{M}_k\) the performance metrics, and \(\mathcal{D}_k\) the final promotion decisions. This formalization ensures that every factor follows an identical, auditable path from conception to selection.

Two principles govern the workflow. First, constrained autonomy: the agent is free to propose new ideas only within explicit scientific constraints, including a fixed variable universe, bounded expression complexity, and strict no-look-ahead rules. Second, evidence accumulation: every round leaves a structured trace, and future proposals are conditioned on past outcomes rather than prompt-level intuition. Together, these principles make the process scalable without giving up empirical discipline. In contrast to conventional prompting pipelines, where the model outputs a static list of factors once, our system learns through repeated interaction with objective performance feedback.

\subsection{Agent Architecture and Functional Modules}
\noindent The architecture contains five operational modules that jointly implement the loop. The first module is a constrained hypothesis generator. It builds candidate factors from interpretable primitives (such as return dynamics, price-relative transforms, volume and liquidity information, and volatility states) under a predefined factor grammar. 
{Specifically, a candidate factor \(f_{i,t}\) for asset \(i\) at time \(t\) is defined as a symbolic composition:
\begin{equation}
f_{i,t} = \mathcal{G} \left( \mathbf{X}_{i,t}, \mathbf{X}_{i,t-1}, \dots, \mathbf{X}_{i,t-k} ; \mathcal{O} \right)
\end{equation}
where \(\mathbf{X}_{i,t}\) denotes the vector of raw primitives (e.g., closing price, trading volume) and \(\mathcal{O}\) is the set of predefined operators (e.g., moving averages, or other technical indicators).}
Operator choices are intentionally transparent, and expression depth is bounded, so factor definitions remain readable and auditable.

The second module is a deterministic execution layer that maps each recipe to a panel-consistent factor series. Cross-sectional transformations are applied within each date, while time-series transformations are applied within each asset history. This strict separation between language-based proposal and code-based computation prevents hidden numerical drift and guarantees reproducibility under identical inputs. The third module is a unified evaluator, which computes a common metric set for every candidate; no factor receives customized scoring logic.

The fourth module is a transparent gatekeeper that converts evaluation outputs into explicit decisions: promote, hold for review, or retire. This avoids unbounded candidate expansion and reallocates research capacity toward empirically promising regions. The fifth module is memory and policy update. It reads structured logs from prior rounds and generates the next candidate set with a balance between exploitation (local variants of survivors) and exploration (new hypotheses from different signal families). This memory-guided diversification is central to reducing search myopia and mitigating factor crowding inside a narrow subspace.

The complete operational flow of this closed-loop discovery is formally summarized in Algorithm \ref{alg:agent_loop}. The specific statistical metrics used in Step 3 and the gatekeeping logic in Step 4 are further detailed in subsequent sections.

\begin{algorithm}[htbp]
\caption{Agentic Closed-Loop Factor Discovery}
\footnotesize
\label{alg:agent_loop}
\begin{algorithmic}[1]
\renewcommand{\algorithmicrequire}{\textbf{Input:}}
\renewcommand{\algorithmicensure}{\textbf{Output:}}

\Require 
    Raw primitives \(\mathbf{X}\), operator set \(\mathcal{O}\), factor grammar \(\mathcal{G}\), evaluation thresholds \(\Theta = \{\tau_{sig}, \tau_{econ}, \tau_{fail}\}\), and in-sample window \(T_{IS}\).
\Ensure 
    Promoted Alpha Library \(\mathcal{L}_{final}\).

\State \textbf{Initialization:} 
\State \(\mathcal{L}_{final} \gets \emptyset\)
\State \(\mathcal{S}_0 \gets \text{Prior research heuristics and financial intuitions}\)

\For{\(k = 1\) \textbf{to} \(K\)}
    \State \textbf{Step 1: Alpha Hypothesis Synthesis} 
    \State Agent generates candidate factors \(\mathcal{F}_k = \{f_1, \dots, f_n\}\) via \(\mathcal{G}\) based on state \(\mathcal{S}_{k-1}\)
    \State \(f_{i,t} = \mathcal{G}(\mathbf{X}_{i,t \dots t-k}; \mathcal{O})\) \Comment{Symbolic formulation of trading signals}

    \State \textbf{Step 2: Factor Construction \& Data Alignment}
    \For{each \(f \in \mathcal{F}_k\)}
        \State Compute raw signal values \(f_{i,t}\) for universe \(\mathcal{N}\) over \(T_{IS}\)
        \State \(\tilde{f}_{i,t} \gets \text{Z-score}(\text{winsorize}(f_{i,t}))\) \Comment{Cross-sectional normalization}
    \EndFor

    \State \textbf{Step 3: Statistical Backtesting \& Metric Estimation} \Comment{Detailed in subsection \ref{subsection:Evaluation}}
    \For{each \(f \in \mathcal{F}_k\)}
        \State Estimate Information Coefficient \(t_{IC}\) and Long-Short Sharpe Ratio \(SR_{LS}\)
        \State \(\mathbf{m}_f \gets [t_{IC}, SR_{LS}, \dots]^\top\) \Comment{Performance vector}
    \EndFor

    \State \textbf{Step 4: Alpha Selection \& Quality Control}
    \For{each \(f \in \mathcal{F}_k\)}
        \If{\(t_{IC} \geq \tau_{sig}\) \textbf{and} \(SR_{LS} \geq \tau_{econ}\)}
            \State \(D(f) \gets \text{Promote to Library}\); \(\mathcal{L}_{final} \gets \mathcal{L}_{final} \cup \{f\}\)
        \ElsIf{\(t_{IC} < \tau_{fail}\)}
            \State \(D(f) \gets \text{Redundant/Discard}\)
        \Else
            \State \(D(f) \gets \text{Re-evaluate/Hold}\)
        \EndIf
    \EndFor

    \State \textbf{Step 5: Research Heuristic Evolution} \Comment{Detailed in subsection \ref{subsection:iterative}}
    \State \(\mathcal{E}_k \gets \{ \mathbf{m}_f \mid f \in \mathcal{F}_k \}\); \(\mathcal{L}_k \gets \{ (f, D(f)) \mid f \in \mathcal{F}_k \}\)
    \State \(\mathcal{S}_k \gets \text{LLM}(\mathcal{S}_{k-1}, \mathcal{E}_k, \mathcal{L}_k)\) \Comment{Update search policy via reflection}
\EndFor

\State \textbf{return} \(\mathcal{L}_{final}\)
\end{algorithmic}
\end{algorithm}

\subsection{Evaluation and Selection Protocol}
\label{subsection:Evaluation}
\noindent Our evaluation protocol is pre-committed before iterative search begins. We use strict time-based separation between in-sample screening and out-of-sample validation, and only in-sample results are allowed to influence promotion decisions. This temporal discipline ensures that policy updates are based on historically available information and prevents contamination from future returns. Because the same protocol is reused across rounds, gains in performance can robustly attributed to the superiority of the generated hypotheses rather than data-snooping via dynamic evaluation rules.

To further control data leakage, we impose information-availability constraints at both the feature and decision layers. At the feature layer, all candidate factors are constructed from contemporaneously observable or lagged inputs only; no forward return, future window statistic, or post-date transformation is allowed in the factor grammar. Cross-sectional preprocessing (such as ranking, z-scoring, and winsorization) is performed date by date\footnote{{To ensure the comparability of factors across different time periods, we apply a cross-sectional Z-score transformation to the raw factor values:
\begin{equation}
\tilde{f}_{i,t} = \frac{\text{winsorize}(f_{i,t}) - \mu_t}{\sigma_t}
\end{equation}
where \(\mu_t\) and \(\sigma_t\) are the cross-sectional mean and standard deviation of the factor values at time \(t\), respectively. The winsorization process caps extreme outliers at the 1st and 99th percentiles to ensure the robustness of the subsequent statistical estimations.}}, so each day is transformed using only same-day cross-sectional information rather than full-sample moments. Time-series operators (such as lags and rolling moments) are applied within each asset history using historical observations up to that date, preventing accidental look-ahead through panel-level operations.
{To formalize the no-look-ahead constraint, we define the information set $\mathcal{F}_t$ available at time $t$. Any candidate factor $f_{i,t}$ must be measurable with respect to this filtration:
\begin{equation}
f_{i,t} \in \mathcal{F}_t = \sigma \left( \{ \mathbf{X}_{j,s} \}_{j \in \mathcal{N}, s \leq t} \right)
\end{equation}
where $\mathcal{N}$ is the universe of assets and $\sigma(\cdot)$ denotes the $\sigma$-algebra generated by historical observations. This ensures that the factor value at $t$ is strictly a function of the past and present, with no dependency on any $s > t$.}

At the decision layer, leakage control is enforced by design: promotion gates are computed exclusively on the in-sample segment, and out-of-sample outcomes are never used to revise thresholds, re-rank candidates, or update the next-round search policy. This separation is critical in an autonomous loop, where repeated iteration can otherwise create implicit test-set feedback. In our workflow, out-of-sample data are used only for final validation and reporting after candidate selection has been fixed.
{The separation between in-sample (IS) and out-of-sample (OOS) periods is defined by the set of dates \( T = T_{IS} \cup T_{OOS} \), where \( \max(T_{IS}) < \min(T_{OOS}) \). The promotion decision \( D \) is a mapping that depends only on the IS history:
\begin{equation}
D(f) = \Psi \left( \{ f_{i,t}, R_{i,t+1} \}_{t \in T_{IS}} \right)
\end{equation}
where \( \Psi \) represents the gatekeeping logic. Thus, the OOS performance
$
\{ f_{i,t}, R_{i,t+1} \}_{t \in T_{OOS}}
$
remains a ``blind'' test set that does not enter the mapping \( \Psi \), thereby preventing the implicit overfitting that often plagues iterative search processes.}

Statistical evaluation is anchored by daily cross-sectional rank IC and its associated t-statistic. Rank IC is preferred as the primary statistical criterion because it directly captures ordering ability and is less sensitive to scaling artifacts. 
{The daily rank IC is computed as the Spearman correlation between the factor scores and the subsequent forward returns:
\begin{equation}
IC_t = \text{Corr} \left( \text{rank}(f_{i,t}), \text{rank}(R_{i,t+1}) \right)
\end{equation}
where \(R_{i,t+1}\) represents the cross-section of asset returns in the next period. The system then aggregates these into a t-statistic, 
\begin{equation}
t_{IC} = \frac{\bar{IC}}{\sigma_{IC} / \sqrt{T}}
\end{equation} 
to ensure statistical significance.}
Economic evaluation is conducted through sorted-portfolio tests. On each date, assets are sorted by factor score into quantiles; we compute quantile returns and a long-short spread (top minus bottom quantile) as the tradable representation of each signal. 
{The return of the long-short portfolio $R_{LS, t}$ is defined as the difference between the mean returns of the top and bottom quantiles:
\begin{equation}
R_{LS, t} = \frac{1}{|Q_{top}|} \sum_{i \in Q_{top}} R_{i,t} - \frac{1}{|Q_{bottom}|} \sum_{i \in Q_{bottom}} R_{i,t}
\end{equation}
where $Q_{top}$ and $Q_{bottom}$ represent the sets of assets in the highest and lowest factor score quantiles, respectively. This spread serves as the basis for calculating the annualized Sharpe ratio\footnote{{The annualized Sharpe ratio of the long-short strategy is then derived from the daily spread returns:
\begin{equation}
\text{SR}_{LS} = \sqrt{252} \cdot \frac{\mathbb{E}[R_{LS, t}]}{\text{Std}(R_{LS, t})}
\end{equation}
where \(\mathbb{E}[R_{LS, t}]\) and \(\text{Std}(R_{LS, t})\) represent the sample mean and standard deviation of the daily long-short returns, respectively.}} and drawdown metrics.}
From these return series, we report annualized return and Sharpe ratio, and we also inspect full quantile profiles to detect non-monotonic patterns.

Selection is rule-based rather than discretionary. A candidate is promoted only if it satisfies minimum statistical thresholds under the fixed protocol. 
{Formally, the gatekeeping decision $D$ for a candidate factor $f$ is determined by a set of pre-defined thresholds $\Theta = \{\tau_{sig}, \tau_{econ}, \tau_{fail}\}$:
\begin{equation}
D(f) = 
\begin{cases} 
\text{Promote} & \text{if } t_{IC} \geq \tau_{sig} \text{ and } \text{SR}_{LS} \geq \tau_{econ} \\
\text{Retire} & \text{if } t_{IC} < \tau_{fail} \\
\text{Hold} & \text{otherwise}
\end{cases}
\end{equation}
where $\tau_{sig}$, $\tau_{econ}$, and $\tau_{fail}$ denote the thresholds for statistical significance, economic viability, and outright failure, respectively. This formalization ensures that only factors exhibiting both robust predictive power and practical profitability enter the final library, while borderline cases are retained for further scrutiny.}
Borderline cases are flagged for targeted robustness checks, and clear failures are retired with explicit annotations. This gatekeeping process serves as the control layer of autonomous discovery: it limits false positives, improves computational efficiency, and creates an auditable rationale for every selection decision.

\subsection{Iterative Learning, Governance, and Extensibility}
\label{subsection:iterative}
\noindent At the end of each round, the system writes a structured experiment record containing candidate definitions, metric outputs, gate outcomes, and next-step actions. This record is not only a logging artifact but also the state representation of the agent. 
{The evolution of the search policy can be formalized as a state-space update:
\begin{equation}
\mathcal{S}_{t+1} = \text{LLM} \left( \mathcal{S}_t, \mathcal{E}_t, \mathcal{L}_t \right)
\end{equation}
where \(\mathcal{S}_t\) is the current knowledge state, \(\mathcal{E}_t\) represents the empirical feedback from the evaluator, and \(\mathcal{L}_t\) is the structured log of failed and successful hypotheses. This ensures that the next generation of factors is conditioned on the cumulative evidence rather than independent random draws.}
The next policy update uses this state to avoid redundant trials, identify successful operator-variable patterns, and deliberately propose diverse alternatives when the current frontier becomes crowded. As a result, the search process behaves as sequential evidence-based learning rather than repeated random generation.

This governance layer also supports replication and post-hoc diagnostics. Because each round is fully documented, researchers can reconstruct how a factor entered the library, which tests it passed, and why competing candidates were rejected. Such traceability is essential for empirical finance applications where model risk and selection bias must be explicitly managed. Finally, the architecture is modular by design: new operators, stricter gate criteria, additional risk controls, or richer transaction-cost models can be integrated without changing the core loop. This modularity allows the framework to adapt across market regimes while retaining methodological continuity.

\begin{figure}[htbp]
\begin{center}
\includegraphics[width=0.99\textwidth]{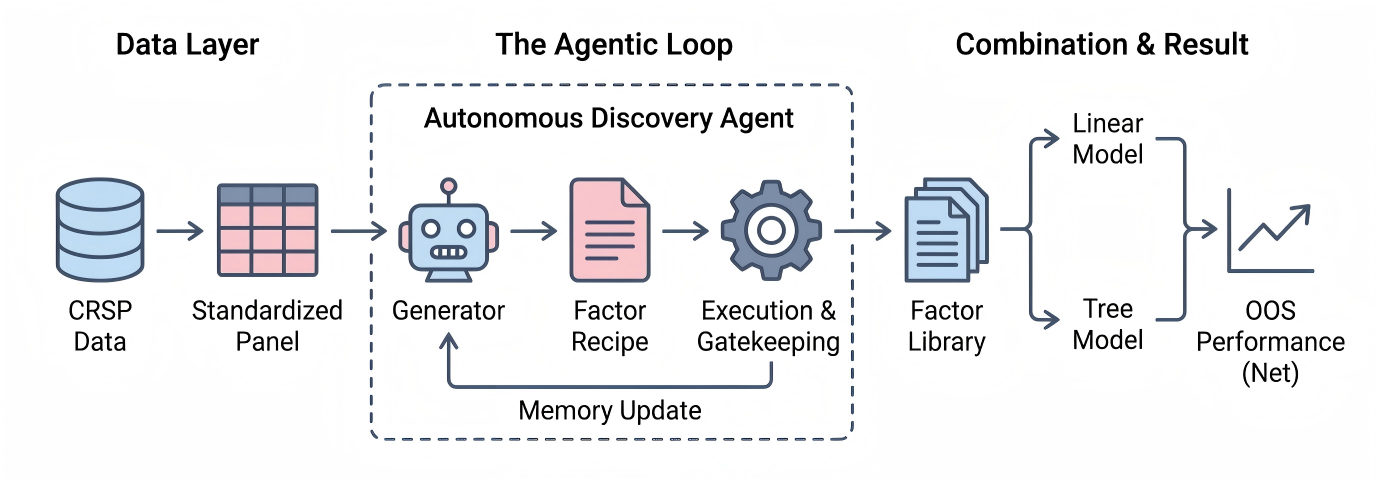}
\caption{\textbf{The Workflow of Agentic AI in Factor Generation}}
\label{fig:workflow}
\end{center}
{
\footnotesize
\noindent
\begin{spacing}{1.4}
This figure illustrates an Agentic AI framework for automated factor generation. The pipeline processes raw data into standardized panels, utilizes an autonomous agent with a memory-update loop to discover valid factors, and applies machine learning models to the resulting factor library to evaluate out-of-sample (OOS) performance.
\end{spacing}
}
\end{figure} 


\section{Data and Factor Overview}
\label{sec:data}
\subsection{Source Data and Sample Construction}
\noindent Our raw equity data are daily stock-level records from the Center for Research in Security Prices, covering January 2004 through December 2024. We construct a unified stock-date panel through a deterministic preprocessing workflow that standardizes identifiers, aligns trading dates, and enforces a consistent structure for downstream factor construction and portfolio evaluation.

The baseline sample design follows standard empirical screening rules for U.S. common equities. 
Following standard data preprocessing conventions in the literature (e.g., \cite{gu2020empirical}), we restrict our universe to common stocks listed on the NYSE, AMEX, and NASDAQ. To mitigate the influence of microcap anomalies and stale-price effects, we further impose a minimum price filter of 5 USD.
In addition, we require at least 252 observations per stock to preserve stability for rolling transformations. The target return is winsorized cross-sectionally at the 1st and 99th percentiles by date before model estimation.

\begin{table}[htbp]
\caption{\textbf{Sample Construction: Daily U.S. Equity Panel, January 2004 to December 2024}}
\par
{\footnotesize
This table summarizes the sequential sample selection process for the daily U.S. equity panel. The sample spans January 2004 through December 2024. Each row reports the remaining observations in millions and unique stocks in thousands after applying the corresponding screen. The filters include exchange eligibility, common share classification, a minimum price threshold of 5 USD, and a minimum trading history of 252 days.
}
\noindent
\begin{center}
{\footnotesize   
    \begin{tabular*}{0.98\linewidth}{@{\extracolsep{\fill}}lrr}
\toprule
Screen & Obs (M) & Stocks (K) \\
\midrule
Raw universe & 39.59 & 20.37 \\
Eligible exchanges & 32.91 & 15.69 \\
Common shares & 21.78 & 10.10 \\
Price $>=$ USD 5 & 16.66 & 9.57 \\
History $>=$ 252 days & 16.51 & 8.05 \\
\bottomrule
\end{tabular*}  
}
\end{center}
\label{tab:sample-construction}
\end{table}

Exhibit~\ref{tab:sample-construction} shows that screening is economically meaningful and not mechanically trivial. The largest contraction occurs at the common-share filter and price filter stages, which is consistent with excluding non-common securities and thinly traded low-price names. Importantly, even after conservative screens, the final panel remains broad, with 8,052 stocks and 16.5 million stock-day observations, providing sufficient cross-sectional depth for daily ranking tests.

For evaluation windows, we use strict time separation to avoid look-ahead bias. The iterative promotion gate is determined using data through December 2020, and the main out-of-sample results reported in the subsequent sections use the full window from January 2021 to December 2024. In addition, we report the post-January 2023 period as a stricter subsample check to assess robustness in a more recent environment.

\begin{figure}[htbp]
\begin{center}
\includegraphics[width=0.75\textwidth]{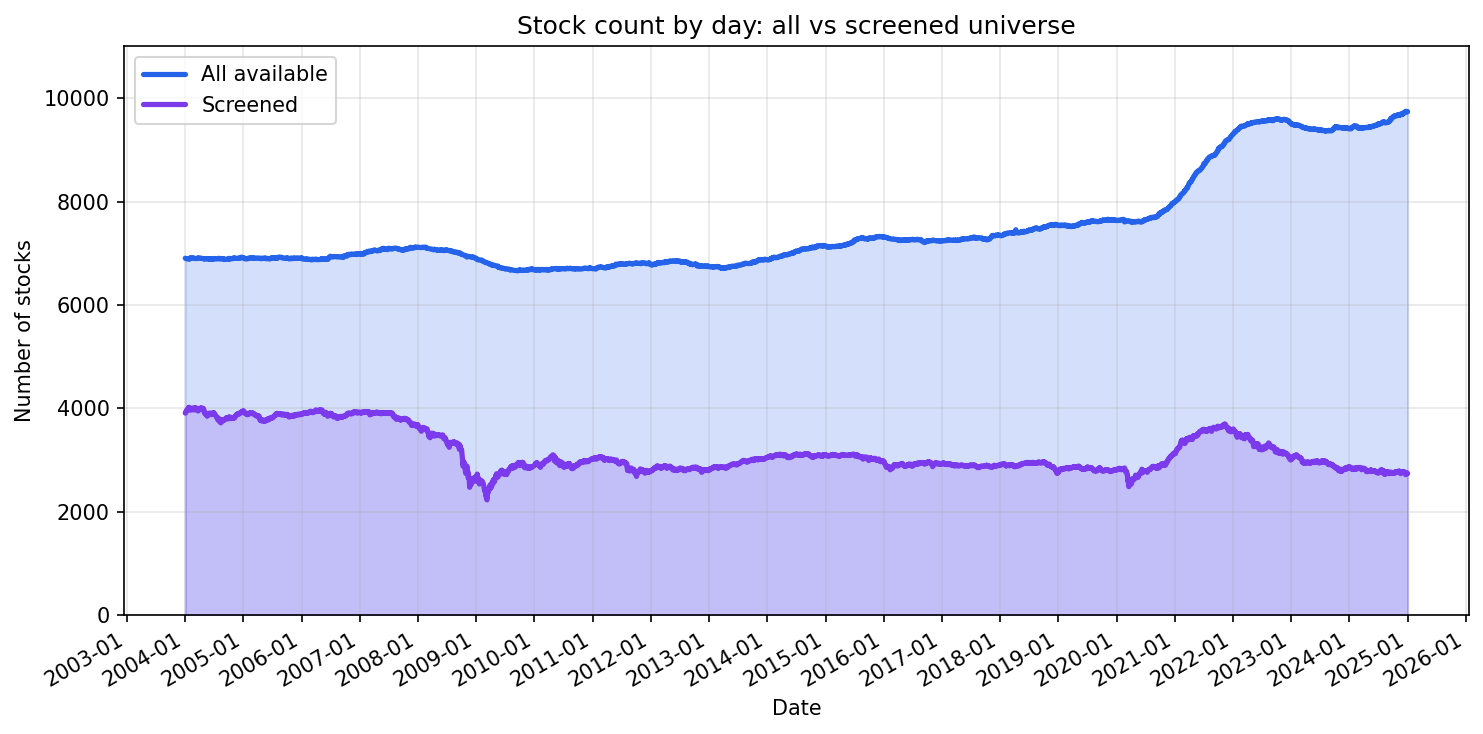}
\caption{\textbf{Evolution of the Stock Universe: Raw CRSP Data vs. Screened Sample}}
\label{fig:data_available_stocks}
\end{center}
{
\footnotesize
\noindent
\begin{spacing}{1.4}
{This figure plots the daily number of stocks available in the raw CRSP database (blue line) and the final sample remaining after applying data screening criteria (purple line). The sample period spans from January 2004 to December 2024. Shaded areas represent the respective universe sizes over time.}
\end{spacing}
}
\end{figure} 

\subsection{Target and Predictor Set}
\noindent The prediction target is the one-day-ahead stock return at the daily frequency. Candidate signals are generated from a compact and interpretable set of stock-native and market-state primitives derived from standard CRSP fields, including stock return, price, trading volume, and broad market return series.

The baseline predictor set includes ten variables: lagged stock return, market return, absolute stock price, trading volume, volume ratio to recent history, 20-day realized volatility, price-to-moving-average ratio, market volatility, volume growth, and a spread proxy when quote data are available. These variables are transformed through transparent operators such as lags, rolling moments, cross-sectional ranks, and arithmetic combinations to form candidate factors.

This constrained design is deliberate. First, it keeps factor definitions auditable and economically interpretable. Second, it limits expression complexity, reducing the risk of overfitting through unconstrained symbolic search. Third, it creates a stable benchmark environment in which improvements can be attributed to better hypotheses and combination logic, rather than to changing data definitions.

\begin{table}[htbp]
\caption{\textbf{Descriptive Statistics of Key Variables, January 2004 to December 2024}}
\par
{\footnotesize
This table reports descriptive statistics for key variables from January 2004 to December 2024. $N$ (M) denotes observations in millions. Returns are reported in percentage points, and share volume is in millions. The statistics include the mean, standard deviation (SD), median, and 99th percentile (P99).
}
\noindent
\begin{center}
{\footnotesize   
    \begin{tabular*}{0.98\linewidth}{@{\extracolsep{\fill}}lcccccc}
\toprule
Variable & Unit & N (M) & Mean & SD & Median & P99 \\
\midrule
Daily stock return & \% & 16.50 & 0.086 & 3.499 & 0.000 & 8.885 \\
Stock price & USD & 16.51 & 118.05 & 5051.21 & 23.12 & 327.95 \\
Share volume & million shares & 16.51 & 1.280 & 5.333 & 0.246 & 16.817 \\
VW market return & \% & 16.51 & 0.043 & 1.150 & 0.080 & 2.980 \\
S\&P return & \% & 16.51 & 0.038 & 1.153 & 0.070 & 2.986 \\
\bottomrule
\end{tabular*} 
}
\end{center}
\label{tab:descriptive-statistics}
\end{table}

Exhibit \ref{tab:descriptive-statistics} indicates several features typical of daily equity panels. First, stock returns exhibit substantial dispersion and heavy tails, motivating robust rank-based evaluation and controlled winsorization of the target variable. Second, both price and volume are strongly right-skewed, with medians far below means, highlighting pronounced firm-size heterogeneity in the cross section. Third, the two market benchmarks display closely aligned first and second moments, indicating that either series provides a stable proxy for aggregate daily market conditions in our predictor set.

\subsection{Data Pipeline and Reproducibility}
\noindent Our empirical pipeline is designed around reproducibility as an identification requirement rather than a software convenience. The complete sequence from raw records to model-ready panels, factor construction, and evaluation outputs is deterministic, so identical inputs generate identical results. This property ensures that cross-round comparisons in the agentic loop reflect changes in hypotheses rather than accidental variation in preprocessing or reporting.

A second principle is information-set consistency. The same variable definitions, sample screens, and transformation rules are maintained across rounds, and the selection stage is always computed on the in-sample segment only. Out-of-sample observations are reserved for validation and are excluded from promotion decisions. Combined with the non-forward factor grammar described above, this structure limits leakage channels and preserves the causal ordering of the research process.

Finally, the framework is fully auditable at the experiment level. Each round records candidate definitions, evaluation statistics, and selection outcomes, allowing the full research path to be reconstructed ex post. This record-keeping makes it possible to verify that reported performance is tied to a specific information set, a fixed protocol, and a transparent sequence of decisions, which is essential for credible empirical claims in systematic factor research.

\section{Empirical Results}
\label{sec:empirics}
\subsection{Single-Factor Portfolios}
\noindent We start the empirical analysis with single-factor portfolio tests as a disciplined benchmark for the broader factor-discovery framework. The key objective at this stage is to establish whether each candidate signal carries economically interpretable and statistically assessable standalone information about next-day cross-sectional stock returns. This benchmark is essential because any subsequent multi-factor improvement is meaningful only if the incremental gains can be evaluated relative to clearly documented single-factor baselines.

This design also serves an identification purpose. By first evaluating one factor at a time under a common portfolio-construction protocol, we separate signal-level content from combination-level engineering and obtain a transparent mapping from factor definition to return-sorting behavior. In this sense, single-factor portfolios provide the minimal empirical unit for understanding where predictability comes from before studying how predictability is combined.
A second motivation is comparability and internal validity. Daily return prediction is particularly exposed to microstructure effects and short-horizon noise. We therefore adopt a common ranking-based evaluation framework for all candidate factors and assess their signal quality under consistent conventions. This design combines rank-correlation metrics with portfolio-sort evidence, so that cross-factor differences can be interpreted as differences in signal content rather than artifacts of changing evaluation rules.
Finally, this section establishes the empirical bridge to the next part of the paper. Once the single-factor baseline is in place, the multi-factor analysis can be framed as a test of incremental value: whether combining signals improves stability and risk-adjusted performance beyond what is visible at the individual-factor level.

\begin{table}[htbp]
\caption{\textbf{Performance Metrics of Agentic AI-Generated Factors}}
\par
{\footnotesize
This table reports multiple performance metrics for single factors generated by our agentic AI framework (see Exhibit \ref{fig:workflow} and Algorithm \ref{alg:agent_loop} for methodological details). The long-short portfolios are constructed by going long on stocks in the top 50\% and shorting those in the bottom 50\% based on factor rankings. Reported metrics include the Sharpe ratio, Information Coefficient (IC), IC Information Ratio (ICIR), Long-only IC (ICL), Long-only ICIR (ICLIR), Sortino ratio, Calmar ratio, annualized returns (Ann. Ret.), and maximum drawdown (Max DD). The factor evaluation period is from 2021.01 to 2024.12. The portfolios are rebalanced at each market day.
}
\noindent
\begin{center}
\scriptsize
\begin{tabular*}{\textwidth}{@{\extracolsep{\fill}}lrrrrrrrrr}
\toprule
 & Sharpe & IC & ICIR & ICL & ICLIR & Sortino & Calmar & Annual Ret & Max DD \\
\midrule
Factor 1 & 2.8593 & 0.0068 & 4.8432 & 0.0033 & 1.0371 & 3.7811 & 3.5915 & 0.1754 & -0.0488 \\
Factor 2 & -0.1767 & 0.0011 & 0.7986 & -0.0022 & -0.6969 & -0.2528 & -0.0898 & -0.0107 & -0.1190 \\
Factor 3 & 2.4140 & 0.0098 & 5.1572 & 0.0061 & 3.4130 & 3.1564 & 2.3044 & 0.2402 & -0.1042 \\
Factor 4 & 0.6535 & 0.0051 & 2.0167 & 0.0003 & 0.1371 & 0.9670 & 0.4057 & 0.0768 & -0.1894 \\
Factor 5 & 1.4069 & 0.0059 & 3.4047 & 0.0003 & 0.1080 & 2.0108 & 1.3124 & 0.1196 & -0.0911 \\
Factor 6 & 2.2597 & 0.0071 & 4.9613 & 0.0036 & 1.5999 & 3.2242 & 1.8383 & 0.1543 & -0.0839 \\
Factor 7 & 0.7182 & -0.0003 & -0.1433 & -0.0000 & -0.0045 & 1.0096 & 0.5023 & 0.0527 & -0.1048 \\
Factor 8 & 1.6628 & 0.0271 & 5.3121 & 0.0131 & 2.6462 & 2.7148 & 1.9476 & 0.3928 & -0.2017 \\
Factor 9 & 1.9421 & 0.0055 & 3.3963 & -0.0008 & -0.3131 & 2.8600 & 2.2886 & 0.1317 & -0.0575 \\
Factor 10 & 1.4424 & 0.0049 & 2.9661 & 0.0004 & 0.1513 & 2.1149 & 1.5943 & 0.1078 & -0.0676 \\
Factor 11 & 0.6228 & 0.0016 & 0.4784 & -0.0000 & -0.0012 & 0.9339 & 0.5539 & 0.0817 & -0.1475 \\
Factor 12 & 0.8362 & 0.0033 & 2.3195 & -0.0002 & -0.1094 & 1.2785 & 0.6246 & 0.0541 & -0.0867 \\
\bottomrule
\end{tabular*}
\end{center}
\label{tab:single long-short}
\end{table}

Exhibit~\ref{tab:single long-short} provides a comprehensive single-factor diagnostic for the agentic AI-generated signals over the 2021.01--2024.12 window. To formally evaluate these signals, we utilize standard performance and risk metrics, including the Information Coefficient (IC), IC Information Ratio (ICIR), Sortino ratio, Maximum Drawdown (MaxDD), and Calmar ratio, which are defined as follows:
\begin{equation}
\begin{aligned}
\text{IC} &= \rho_{\text{rank}}(f_t, R_{t+1}) \\
\text{ICIR} &= \frac{\mu_{\text{IC}}}{\sigma_{\text{IC}}} \\
\text{Sortino} &= \frac{R_p - R_f}{\sigma_d} \\
\text{MaxDD} &= \max_{\tau \in (0, T)} \frac{\max_{t \in (0, \tau)} P_t - P_\tau}{\max_{t \in (0, \tau)} P_t} \\
\text{Calmar} &= \frac{R_p}{\text{MaxDD}}
\end{aligned}
\end{equation}
Exhibit \ref{tab:single long-short} shows that standalone signal quality is heterogeneous rather than uniformly strong. Several factors exhibit attractive long-short profiles with strong Sharpe ratios and economically meaningful annualized returns, while the highest-IC factor (Factor 8) also stands out on ranking ability. By contrast, several other factors appear materially weaker out of sample, including at least one case (Factor 2) with poor portfolio monetization despite a small positive IC. This pattern suggests that the agentic framework is capable of generating genuinely useful signals, but it does not imply that every promoted factor is equally investable on a standalone basis. The heterogeneity in IC, Sharpe, and drawdown profiles motivates the multi-factor aggregation analysis that follows.

\begin{landscape}
\begin{table}[htbp]
\caption{\textbf{Univariate Portfolio Sorts: Decile Returns of Agentic AI-Generated Factors}}
\label{tab:factor_by_group}
\par
{\footnotesize
This table presents the average equal-weighted returns for decile portfolios sorted by factors discovered by our agentic AI framework. At each rebalancing date, stocks are ranked by their respective factor scores and partitioned into ten portfolios. ``Low'' and ``High'' denote the bottom and top deciles, respectively. The ``High$-$Low'' column reports the return spread between the top and bottom deciles. All returns are expressed in decimal form. The portfolios are rebalanced at each market day.
}
\noindent
\begin{center}
\footnotesize
\begin{tabular*}{\linewidth}{@{\extracolsep{\fill}}lrrrrrrrrrrr}
\toprule
& \multicolumn{10}{c}{Deciles} & \\
\cmidrule(lr){2-11}
Factor & Low & 2 & 3 & 4 & 5 & 6 & 7 & 8 & 9 & High & High-Low \\
\midrule
Factor 1 & 7.6080 & 9.3167 & 11.5033 & 9.1529 & 12.3174 & 12.6893 & 13.7833 & 12.6944 & 13.9916 & 23.9337 & 16.3258 \\
         & (1.1435) & (0.9004) & (1.0936) & (0.8718) & (1.1946) & (1.2270) & (1.3190) & (1.2090) & (1.3539) & (3.5077) & (5.7073) \\
Factor 2 & 12.7183 & 10.5950 & 12.0842 & 12.7672 & 11.3010 & 13.7137 & 12.8309 & 15.3788 & 13.8296 & 11.7832 & -0.9351 \\
         & (1.9314) & (1.0272) & (1.1426) & (1.2142) & (1.0916) & (1.3273) & (1.2288) & (1.4678) & (1.3441) & (1.7285) & (-0.3526) \\
Factor 3 & 5.5370 & 7.4050 & 9.4599 & 11.6211 & 11.5141 & 9.7888 & 12.5611 & 15.7002 & 15.9265 & 27.4833 & 21.9463 \\
         & (0.8371) & (0.7320) & (0.9123) & (1.2202) & (1.1812) & (0.9453) & (1.2198) & (1.5242) & (1.5420) & (3.0366) & (4.8185) \\
Factor 4 & 10.8439 & 10.0526 & 9.0151 & 11.5517 & 11.7319 & 14.1782 & 13.1807 & 11.5061 & 15.9148 & 19.0283 & 8.1844 \\
         & (2.0288) & (1.0472) & (0.8604) & (1.0963) & (1.1223) & (1.3544) & (1.2762) & (1.1177) & (1.5488) & (1.9991) & (1.3043) \\
Factor 5 & 8.8644 & 8.3713 & 7.8395 & 11.2448 & 11.8396 & 12.5005 & 14.1687 & 14.2334 & 17.4292 & 20.5067 & 11.6424 \\
         & (1.3531) & (0.8181) & (0.7503) & (1.0800) & (1.1480) & (1.2306) & (1.3855) & (1.4026) & (1.7279) & (2.4144) & (2.8083) \\
Factor 6 & 6.5800 & 9.1634 & 11.6692 & 11.6242 & 13.0579 & 12.3224 & 11.9336 & 12.3290 & 17.1525 & 21.1415 & 14.5615 \\
         & (0.7852) & (1.0063) & (1.1449) & (1.1298) & (1.2778) & (1.2132) & (1.1613) & (1.2157) & (1.8851) & (2.4920) & (4.5105) \\
Factor 7 & 12.5296 & 11.5006 & 10.8795 & 12.0201 & 11.0813 & 12.0757 & 11.4794 & 13.8562 & 13.6509 & 17.9456 & 5.4160 \\
         & (1.4147) & (1.1236) & (1.0538) & (1.1733) & (1.0781) & (1.1758) & (1.1434) & (1.3972) & (1.4839) & (2.5567) & (1.4335) \\
Factor 8 & -13.7849 & -0.1989 & 7.2228 & 12.6352 & 22.0386 & 21.0944 & 17.6189 & 19.6974 & 19.2048 & 21.6353 & 35.4202 \\
         & (-1.0195) & (-0.0171) & (0.6602) & (1.1520) & (1.9228) & (2.0029) & (1.9113) & (2.4984) & (2.8172) & (3.3056) & (3.3190) \\
Factor 9 & 7.7250 & 5.7652 & 9.8208 & 12.3451 & 12.0076 & 13.8103 & 14.2841 & 15.4949 & 15.4298 & 20.3107 & 12.5857 \\
         & (1.0693) & (0.5589) & (0.9452) & (1.2059) & (1.1722) & (1.3628) & (1.4140) & (1.5359) & (1.5458) & (2.4836) & (3.8764) \\
Factor 10 & 8.6473 & 8.2489 & 10.1795 & 11.2505 & 12.5063 & 12.9636 & 12.1496 & 14.9647 & 16.9427 & 19.1523 & 10.5051 \\
          & (1.2554) & (0.8022) & (0.9705) & (1.0942) & (1.2188) & (1.2692) & (1.2009) & (1.4807) & (1.6848) & (2.3136) & (2.8791) \\
Factor 11 & 8.2225 & 9.1330 & 11.2494 & 10.7664 & 13.8119 & 13.5295 & 11.0894 & 13.7121 & 18.3941 & 17.0931 & 8.8706 \\
          & (0.8459) & (0.8546) & (1.0584) & (1.0137) & (1.2979) & (1.2584) & (1.0298) & (1.3099) & (2.0924) & (3.9138) & (1.2432) \\
Factor 12 & 11.4317 & 9.9040 & 10.9462 & 11.5047 & 12.2840 & 12.8936 & 12.0737 & 14.1727 & 14.8922 & 16.9192 & 5.4875 \\
          & (1.3371) & (1.0853) & (1.0853) & (1.1321) & (1.2090) & (1.2800) & (1.1864) & (1.4098) & (1.6299) & (1.9329) & (1.6691) \\
\bottomrule
\end{tabular*}
\end{center}
\end{table}
\end{landscape}

Exhibit \ref{tab:factor_by_group} provides a cross-sectional portfolio-sort view of each factor and reveals meaningful but uneven sorting ability across the signal set. Several factors generate economically large and statistically credible High–Low spreads, while monotonicity is not universal across all candidates and a subset of factors shows weak or statistically fragile spread returns, including at least one case with a negative High–Low spread. The evidence therefore supports a selective interpretation: the framework discovers multiple useful ranking signals, but their standalone sorting quality varies enough to justify subsequent multi-factor integration.

\begin{table}[htbp]
\caption{\textbf{Risk-Adjusted Alphas of Agentic AI-Generated Factors}}
\label{tab:factor_alpha_oos}
\par
{\footnotesize
This table presents annualized out-of-sample alphas and corresponding $t$-statistics for factors discovered by our agentic AI framework. Alphas are estimated via time-series regressions of long-short portfolio returns on the CAPM \citep{sharpe1964capital}, FF3 \citep{fama1993common}, FF5 \citep{fama2015five}, and FF6 models (FF5 plus momentum factor proposed by \cite{jegadeesh1993returns}). The long-short portfolios are constructed by sorting stocks into deciles based on factor rankings. All alphas are reported in percentages, and $t$-statistics (in parentheses) are adjusted for heteroskedasticity and autocorrelation following \cite{newey1987hypothesis}.
}
\begin{center}
\footnotesize
\begin{tabular*}{\linewidth}{@{\extracolsep{\fill}}lcccc}
\toprule
Factor & CAPM \(\alpha\) & FF3 \(\alpha\) & FF5 \(\alpha\) & FF6 \(\alpha\) \\
\midrule
Factor 1 & 0.134 & 0.137 & 0.138 & 0.138 \\
         & (4.57) & (4.71) & (4.74) & (4.74) \\
Factor 2 & -0.037 & -0.038 & -0.040 & -0.040 \\
         & (-1.30) & (-1.34) & (-1.43) & (-1.43) \\
Factor 3 & 0.190 & 0.196 & 0.195 & 0.196 \\
         & (4.22) & (4.30) & (4.33) & (4.37) \\
Factor 4 & 0.054 & 0.058 & 0.056 & 0.057 \\
         & (0.88) & (0.95) & (0.93) & (0.95) \\
Factor 5 & 0.088 & 0.089 & 0.088 & 0.088 \\
         & (2.18) & (2.21) & (2.18) & (2.23) \\
Factor 6 & 0.115 & 0.118 & 0.115 & 0.115 \\
         & (3.61) & (3.70) & (3.64) & (3.65) \\
Factor 7 & 0.023 & 0.021 & 0.021 & 0.022 \\
         & (0.62) & (0.57) & (0.57) & (0.59) \\
Factor 8 & 0.331 & 0.320 & 0.315 & 0.318 \\
         & (3.32) & (3.26) & (3.24) & (3.26) \\
Factor 9 & 0.098 & 0.099 & 0.100 & 0.100 \\
         & (3.08) & (3.11) & (3.17) & (3.21) \\
Factor 10 & 0.077 & 0.077 & 0.077 & 0.078 \\
          & (2.14) & (2.15) & (2.18) & (2.23) \\
Factor 11 & 0.060 & 0.052 & 0.053 & 0.053 \\
          & (0.90) & (0.77) & (0.80) & (0.80) \\
Factor 12 & 0.028 & 0.026 & 0.021 & 0.022 \\
          & (0.83) & (0.78) & (0.65) & (0.67) \\
\bottomrule
\end{tabular*}
\end{center}
\end{table}

Exhibit \ref{tab:factor_alpha_oos} evaluates whether the single-factor portfolios survive standard risk adjustment. The evidence is concentrated rather than universal. The stronger factors retain positive and statistically meaningful alpha across the benchmark models, while several weaker candidates lose significance after controlling for standard market and style exposures. The stability of the stronger alphas across CAPM, FF3, FF5, and FF6 nevertheless suggests that the best agent-generated signals are not merely repackaging canonical equity factors.

Overall, the single-factor evidence supports three conclusions. First, the agentic pipeline can identify factors with economically meaningful and statistically credible spread returns, but the quality of those factors is heterogeneous. Second, the stronger signals retain alpha under standard risk adjustment, which suggests that they contain information beyond canonical factor exposures. Third, precisely because standalone performance varies across candidates, disciplined multi-factor integration is essential for improving aggregate out-of-sample performance.

\subsection{Multivariate combination}

\noindent Having established the standalone properties of individual signals, we next evaluate whether combining agentic AI-generated factors can produce more stable and economically meaningful out-of-sample performance. The motivation is straightforward: single-factor evidence is informative but heterogeneous, and practical portfolio construction typically relies on aggregation to reduce idiosyncratic noise, diversify factor-specific risks, and improve implementation robustness. In this subsection, we therefore move from univariate sorting diagnostics to integrated portfolio formation, and test whether a disciplined multi-factor design can deliver incremental value beyond what is attainable from any single signal in isolation.

\begin{table}[htbp]
\caption{\textbf{Out-of-Sample Performance of the Composite Long-Short Strategy}}\par

{This table reports out-of-sample performance of the (linear) composite long-short strategy. Panel A summarizes full-window gross results for 2021 January--2024 December. Panel B reports gross performance by calendar quarter to assess temporal stability. Ann. Return and Ann. Vol. denote annualized return and annualized volatility. Max DD denotes maximum drawdown. The portfolios are rebalanced at each market day.
}
\noindent
\begin{center}
{\footnotesize

\begin{tabular*}{\textwidth}{@{\extracolsep{\fill}}lrrrrrr}
\toprule
\multicolumn{1}{l}{} & Period Ret. (\%) & Ann. Ret. (\%) & Ann. Vol. (\%) & Sharpe & Max DD (\%) & N \\
\midrule
\multicolumn{7}{c}{\textbf{Panel A: OOS 2021 January-2024 December}} \\
\midrule
Long-Short & 470.43 & 54.81 & 16.40 & 2.75 & -13.41 & 1004 \\
\midrule
\multicolumn{7}{c}{\textbf{Panel B: Quarterly gross}} \\
\midrule
2021Q1 & 9.68 & 46.49 & 22.65 & 1.80 & -6.18 & 61 \\
2021Q2 & 12.61 & 60.81 & 15.22 & 3.20 & -6.90 & 63 \\
2021Q3 & 12.04 & 56.45 & 12.38 & 3.68 & -5.88 & 64 \\
2021Q4 & 24.00 & 133.29 & 12.43 & 6.89 & -2.67 & 64 \\
2022Q1 & 21.25 & 118.82 & 22.03 & 3.67 & -5.80 & 62 \\
2022Q2 & 18.38 & 98.53 & 22.20 & 3.20 & -6.46 & 62 \\
2022Q3 & 1.42 & 5.71 & 23.13 & 0.35 & -10.70 & 64 \\
2022Q4 & 14.08 & 69.39 & 19.94 & 2.74 & -4.40 & 63 \\
2023Q1 & 1.37 & 5.69 & 17.88 & 0.40 & -10.04 & 62 \\
2023Q2 & 3.03 & 12.89 & 10.86 & 1.17 & -5.08 & 62 \\
2023Q3 & 18.04 & 94.17 & 10.92 & 6.14 & -2.52 & 63 \\
2023Q4 & 4.76 & 20.45 & 14.93 & 1.32 & -9.70 & 63 \\
2024Q1 & 5.92 & 26.84 & 11.11 & 2.20 & -3.49 & 61 \\
2024Q2 & 17.29 & 89.25 & 10.35 & 6.22 & -1.88 & 63 \\
2024Q3 & 17.85 & 90.92 & 15.37 & 4.29 & -5.97 & 64 \\
2024Q4 & 5.85 & 25.55 & 11.26 & 2.08 & -2.89 & 63 \\
\bottomrule
\end{tabular*}
}
\end{center}
\label{tab:oos-headline}
\end{table}

While individual factors demonstrate standalone predictive power, they may capture overlapping market dynamics. The purpose of aggregation is to exploit their orthogonal information. To achieve this, we employ both a linear model (to capture independent additive premiums) and a tree-based LightGBM model (to capture complex, conditional interactions among these signals) for complementary reasons. The linear specification provides a transparent benchmark with low estimation variance and directly interpretable factor loadings, which is useful for identifying broad directional contributions of each signal. The tree-based specification is introduced to capture potential nonlinearities, interaction effects, and regime-dependent thresholds that are difficult to represent in a purely additive linear structure. Evaluating both models under the same train--test protocol allows us to distinguish robust information that survives across model classes from gains that are model-specific.

Exhibit \ref{tab:oos-headline} reports strong gross out-of-sample performance for the linear composite long-short strategy over the main 2021–2024 evaluation window. The strategy delivers a gross annualized return of 54.81\%, annualized volatility of 16.40\%, and a Sharpe ratio of 2.75. The quarterly breakdown suggests that performance is not concentrated in a single short episode, as the reported gross return remains positive throughout the sample. At the same time, quarterly annualized-return figures should be interpreted with caution because annualization mechanically amplifies short-window outcomes. Overall, the evidence indicates that combining the discovered signals materially improves stability and tradability relative to the more uneven standalone factor results.

\begin{table}[htbp]
\caption{\textbf{Risk-adjusted alpha (annualized) and $t$-statistics: combination portfolios}}
\par
{\footnotesize
{The table presents equal-weighted portfolio returns sorted by composite factor scores. We present the classic CAPM \citep{sharpe1964capital} alpha with \cite{newey1987hypothesis}'s adjusted t-statistics in parentheses below the corresponding alphas. The three- and five-factor models correspond to the classical factor models by \cite{fama1993common,fama2015five}} 
}
\noindent
\begin{center}
\footnotesize
\begin{tabular*}{\linewidth}{@{\extracolsep{\fill}}lcccc}
\toprule
Portfolio & CAPM $\alpha$ & FF3 $\alpha$ & FF5 $\alpha$ & FF6 $\alpha$ \\
\midrule
Linear long-short & 0.425 & 0.417 & 0.412 & 0.414 \\
                  & (5.334) & (5.304) & (5.315) & (5.324) \\
Linear long-only  & 0.293 & 0.302 & 0.300 & 0.300 \\
                  & (4.170) & (4.327) & (4.340) & (4.338) \\
LGBM long-short   & 0.315 & 0.311 & 0.310 & 0.311 \\
                  & (6.986) & (6.871) & (6.917) & (6.904) \\
LGBM long-only    & 0.240 & 0.251 & 0.251 & 0.250 \\
                  & (3.248) & (3.438) & (3.451) & (3.441) \\
\bottomrule
\end{tabular*}
\end{center}
\label{tab:combination_alpha_oos}
\end{table}

Exhibit \ref{tab:combination_alpha_oos} provides strong evidence that the combination portfolios retain substantial abnormal performance after standard risk adjustment. Across CAPM, FF3, FF5, and FF6, all reported portfolio constructions deliver positive and statistically significant annualized alphas. For the linear long-short portfolio, the alpha remains stable as the benchmark model becomes richer, ranging from 0.425 ($t=5.33$) under CAPM to 0.414 ($t=5.32$) under FF6. The nonlinear LightGBM specification has a lower point estimate (e.g., 0.311 under FF6), but its t-statistics remain consistently high across specifications (around 6.90). Taken together, these results suggest that the multi-factor signals are not simply repackaging exposures to standard equity risk factors.

\begin{figure}[htbp]
\begin{center}
\includegraphics[width=0.99\textwidth]{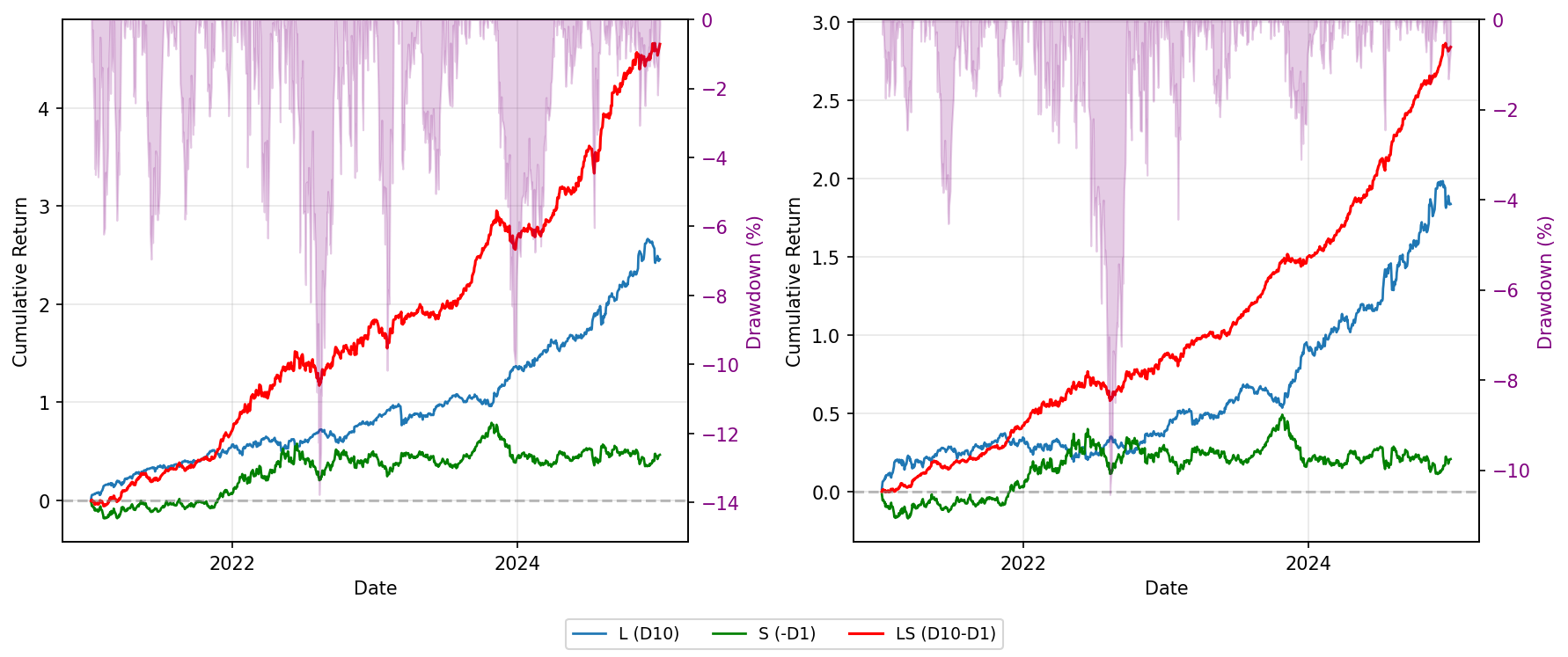}
\caption{\textbf{Cumulative Returns of Multi-factor Portfolios: Linear vs. LightGBM Factor Aggregation}}
\label{fig:multivariate_LandS_cum}
\end{center}
{
\footnotesize
\noindent
\begin{spacing}{1.4}
This figure displays the out-of-sample cumulative returns of portfolios constructed by aggregating multiple Agentic AI-generated factors. The left panel illustrates the performance using a linear aggregation model, while the right panel shows the results using a LightGBM (LGBM) model. Each plot depicts four series: the long-only portfolio (Decile 10), the short-only portfolio (Decile 1), the resulting long-short portfolio (High-Low), and the maximum drawdown of the long-short portfolio (numbers in percentage \%).
\end{spacing}
}
\end{figure} 

Exhibit~\ref{fig:multivariate_LandS_cum} compares the out-of-sample cumulative return paths under two aggregation schemes. Two patterns are central. First, both aggregation methods generate persistent upward trends in the long and short legs, indicating that the discovered factor set remains monetizable after combination. Second, the spread portfolio remains broadly positive over the sample, but its trajectory differs across models: the linear specification exhibits larger interim drawdown episodes, whereas LightGBM shows a smoother drawdown profile and faster stabilization after stress periods. This is consistent with nonlinear aggregation better capturing interaction effects among factors.

\begin{figure}[htbp]
\begin{center}
\includegraphics[width=0.99\textwidth]{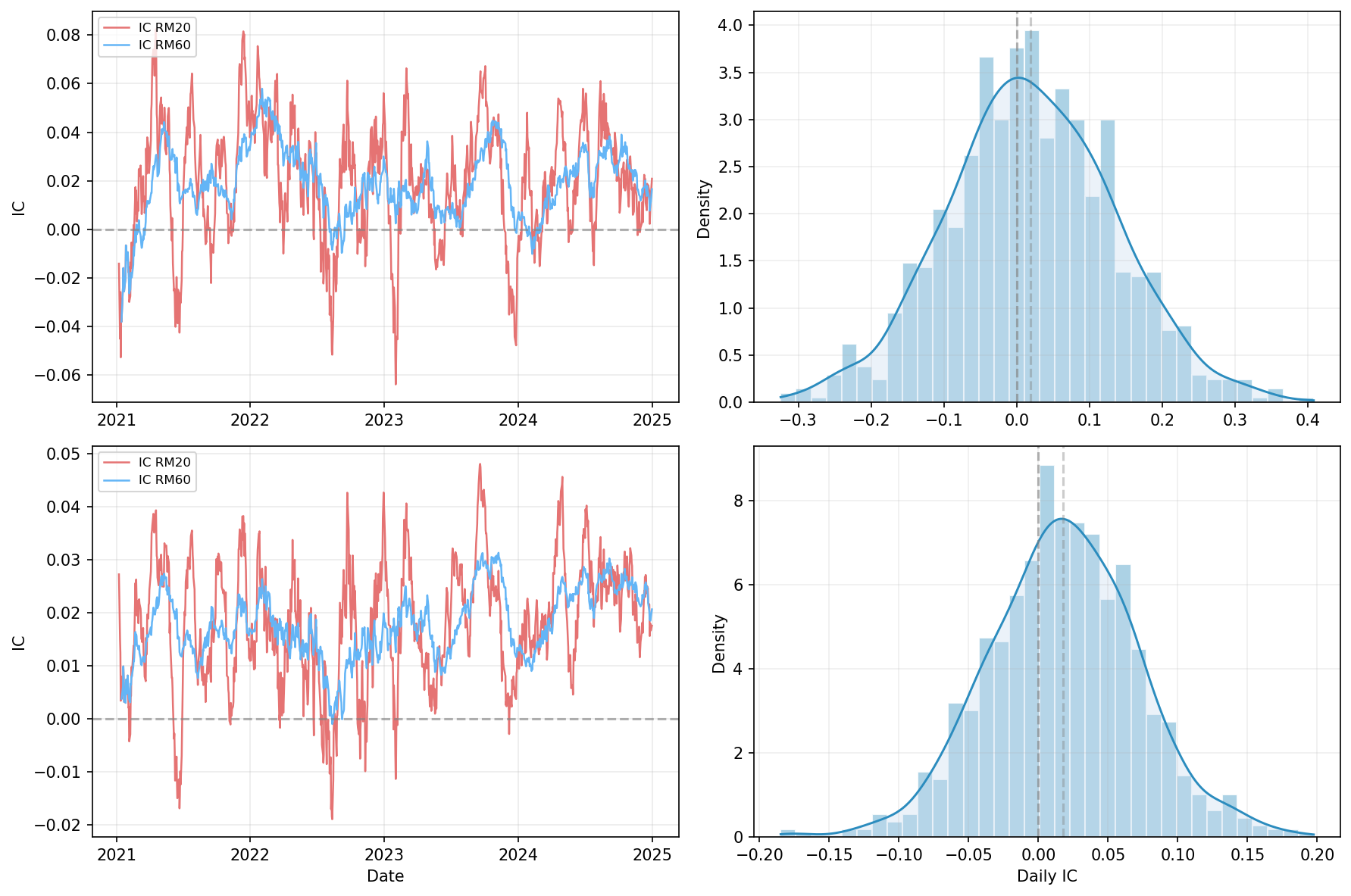}
\caption{\textbf{Information Coefficient (IC) Diagnostics: Linear vs. LightGBM Aggregation}}
\label{fig:multivariate_ic}
\end{center}
{
\footnotesize
\noindent
\begin{spacing}{1.4}
This figure presents the diagnostic analysis of daily rank Information Coefficients (IC) for the composite signals. The upper panel displays the IC metrics for the linear aggregation model, while the lower panel shows the results for the LightGBM (LGBM) aggregation model. The left plot illustrates the daily IC time series with rolling averages, and the right plot displays the empirical distribution of the daily IC values.
\end{spacing}
}
\end{figure} 

Exhibit~\ref{fig:multivariate_ic} compares IC dynamics and provides mechanism-level support for the portfolio outcomes. The linear model shows ICs fluctuating around zero, whereas the LightGBM model maintains a systematically higher IC level over most of the sample, with its rolling means remaining predominantly above zero. The distributional panels reinforce this, showing the LightGBM daily IC distribution shifted to the right with a thicker positive tail. This implies that nonlinear aggregation improves both the level and persistence of predictive information.

Exhibit \ref{tab:decile-performance} reports the out-of-sample performance of decile portfolios. The results reveal a broadly monotonic relationship for the composite signal: the cumulative period return increases from -47.01\% for D1 to 245.72\% for D10, while the annualized Sharpe ratio rises from -0.505 to 2.313. This confirms the model's cross-sectional predictive power. The D10-D1 long-short spread yields a cumulative gross return of 465.07\% with an annualized Sharpe ratio of 2.715. After applying a one-way turnover cost of 3 bps, the corresponding net long-short spread remains economically meaningful, with a cumulative return of 305.57\% and an annualized Sharpe ratio of 2.211.

\begin{table}[htbp]
\caption{\textbf{Decile Portfolio Performance in the Out-of-Sample Window}}\par
{\footnotesize
This table reports decile-level performance for score-sorted portfolios and the D10-D1 long-short spread. Panel A reports decile portfolios. Panel B reports long-short performance. The Gross row reports the long-short spread (D10-D1) using gross daily returns. The Net row deducts a one-way transaction cost of 3 bps from daily turnover: $c_t = 0.0003 \times \mathrm{Turnover}_t$, and $r^{\mathrm{net}}_t = r^{\mathrm{gross}}_t - c_t$. $\mathrm{Turnover}_t$ is one-way daily long-short turnover in decimal form (1.0 = 100\%). Period Return is $\prod_t (1+r_t)-1$, and Ann. Sharpe is $\sqrt{252}\,\bar r/\sigma(r)$ based on daily returns.
}
\noindent
\begin{center}
{\footnotesize
\begin{tabular*}{\textwidth}{@{\extracolsep{\fill}}lrrr}
\toprule
Portfolio & Period Return (\%) & Ann. Sharpe & N Days \\
\midrule
\multicolumn{4}{c}{\textbf{Panel A: Decile (gross)}} \\
\midrule
D1 & -47.01 & -0.505 & 1004 \\
D2 & 6.14 & 0.180 & 1004 \\
D3 & 35.72 & 0.466 & 1004 \\
D4 & 38.48 & 0.507 & 1004 \\
D5 & 48.75 & 0.616 & 1004 \\
D6 & 65.33 & 0.777 & 1004 \\
D7 & 83.99 & 0.934 & 1004 \\
D8 & 87.94 & 0.973 & 1004 \\
D9 & 133.03 & 1.290 & 1004 \\
D10 & 245.72 & 2.313 & 1004 \\
\midrule
\multicolumn{4}{c}{\textbf{Panel B: Long-short spread (D10-D1)}} \\
\midrule
Gross & 465.07 & 2.715 & 1004 \\
Net & 305.57 & 2.211 & 1004 \\
\bottomrule
\end{tabular*}
}
\end{center}
\label{tab:decile-performance}
\end{table}

\begin{figure}[htbp]
\begin{center}
\includegraphics[width=0.75\textwidth]{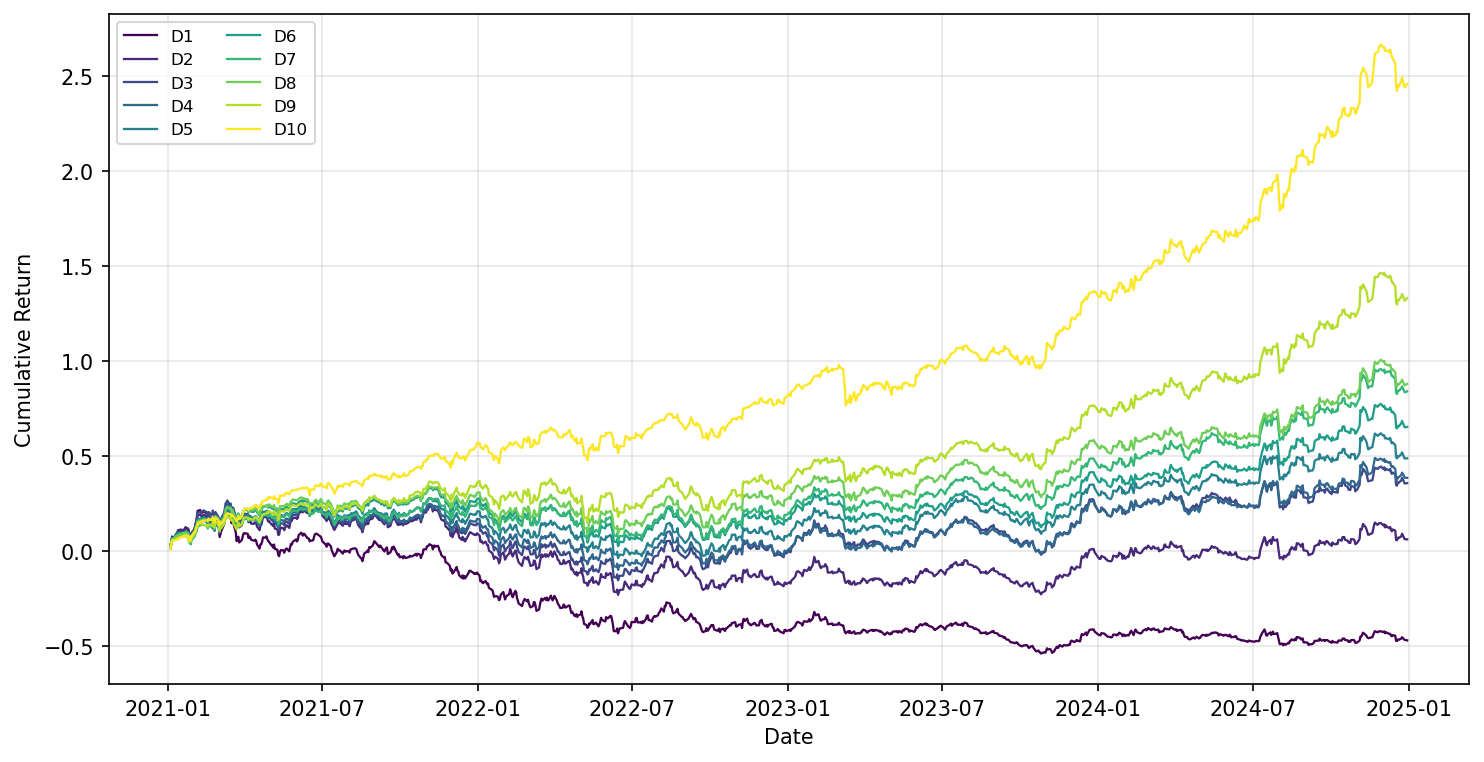}
\caption{\textbf{Cumulative Performance of Decile Portfolios Sorted by Aggregated Signal}}
\label{fig:quantile-monotonicity}
\end{center}
{
\footnotesize
\noindent
\begin{spacing}{1.4}
This figure displays the cumulative returns of ten decile portfolios (D1 to D10) constructed by sorting assets based on their aggregated factor scores. The performance is tracked over the out-of-sample period from 2021 to 2024.
\end{spacing}
}
\end{figure}

Exhibit \ref{fig:quantile-monotonicity} illustrates the cumulative returns of the decile portfolios, revealing a striking ``fan-out'' effect. The top decile (D10) exhibits a consistent upward trajectory, while the bottom decile (D1) shows a persistent decline. This clear separation throughout the 1004-day out-of-sample period demonstrates encouraging cross-sectional discrimination and signal stability, validating the effectiveness of the factor aggregation methodology.

\section{Mitigation of Data-Mining Concerns and Economic Rationale}
\label{sec:mitigation}
\noindent Given the substantial out-of-sample alpha reported in Section \ref{sec:empirics}, a natural concern is whether such performance stems from extensive data-mining or genuine economic regularities. A central challenge in automated factor discovery is data-mining bias---the tendency to uncover spurious patterns that lack economic substance and fail to persist out-of-sample \citep{harvey2016cross}. In the era of the ``Factor Zoo,'' the risk of $p$-hacking is particularly acute for AI-driven methods.
This section validates the integrity of our discovery process through two complementary lenses. First, we detail the stringent statistical hurdles, multi-objective filters, and strict temporal isolation protocols designed to suppress spurious discoveries. Second, we move beyond numerical performance to analyze the ``anatomy'' of the agent-generated signals, demonstrating that the framework's output aligns with established market microstructure and behavioral finance theories. By synthesizing statistical rigor with economic interpretability, we ensure that the reported alpha represents structural market insights rather than transient noise.

\subsection{Economic Regularization and the ``Lucky Factor'' Filter}
\noindent 
A key concern in autonomous factor discovery is the identification of lucky factors—signals that appear significant due to extensive search rather than genuine economic mechanisms \citep{harvey2016cross}. The proliferation of these spurious signals is not merely a statistical artifact, but is deeply rooted in human behavioral biases. As conceptualized in the scientific outlook of financial economics \citep{harvey2017presidential}, human cognition is evolutionarily wired to tolerate high Type I errors (false positives). In a primal environment, the energetic cost of reacting to a false alarm (e.g., fleeing from rustling grass that turns out to be wind) is negligible compared to the fatal cost of a Type II error (e.g., ignoring the rustling grass that conceals a predator). In empirical finance, this asymmetric survival heuristic translates into a pervasive psychological urge to over-interpret random noise as profitable patterns, inevitably driving human researchers toward data mining and the expansion of the ``factor zoo.''

Our Agentic AI framework provides a structural safeguard against this human cognitive flaw. Unlike human researchers who are prone to HARKing (Hypothesizing After the Results are Known)—where a high Sharpe ratio is retrospectively used to justify a spurious narrative—our framework enforces a strict deductive process. By leveraging the vast prior knowledge embedded in Large Language Models, the agent operates within a semantic search space rather than a purely mathematical one. To mitigate the risk of spurious discoveries, our framework embeds economic interpretability directly into the factor generation process. Specifically, the agent generates candidate factors under explicit interpretability constraints. The mathematical expression of a factor and its economic rationale are produced jointly rather than sequentially. This ensures that each proposed signal is accompanied by a plausible economic explanation for why the transformation of price–volume information may predict future returns, fundamentally constraining the brute-force data mining that plagues traditional quantitative research.

This design effectively introduces economic regularization on the symbolic search space. 
Unlike traditional automated machine learning methods (such as genetic programming) that conduct unconstrained random walks across a vast mathematical landscape \citep{neely1997technical,brogaard2023machine}\footnote{Specifically, arbitrarily combining mathematical functions and operators often generates economically meaningless rules, which merely inflates the computational cost of the search process. This is largely because the size of the search space is astronomically large due to the multitude of possible mathematical combinations.}, our Agentic AI leverages the prior financial knowledge embedded in its pre-trained weights. 
By restricting the admissible set of transformations to economically interpretable ones, the framework reduces the likelihood of discovering spurious patterns that arise purely from combinatorial exploration of mathematical operators. 
In essence, the requirement of an explicit economic rationale acts as a Bayesian prior that drastically shrinks the hypothesis search space to an economically meaningful subspace. 
This fundamentally reduces the multiple-testing burden before any statistical hurdles are even applied. 
As a result, the agent prioritizes signals that are both statistically predictive and economically meaningful, filtering out a large class of purely data-mined candidates \citep{harvey2020false}.

\subsection{Strict Statistical Hurdles and Multiple Testing Adjustment}
\noindent To account for the thousands of implicit regressions performed during the discovery process, we eschew the traditional $t > 2.0$ threshold. Instead, we adopt the more stringent significance level recommended by \cite{harvey2016cross} for the current research era:
\begin{equation}
|t\text{-statistic}| > 3.0
\end{equation}
This high hurdle is designed to control the Type I error rate (false discoveries) in a multiple testing environment. By requiring a $t$-stat of 3.0, we ensure that the discovered factors represent genuine market anomalies rather than transient data-mining artifacts. Furthermore, we evaluate the factor's contribution to reducing pricing errors, a concept aligned with the \textit{Scaled Intercept} (SI) metric in \cite{harvey2021lucky}, ensuring that new factors provide incremental explanatory power relative to established benchmarks.

\subsection{Multi-Dimensional Robustness Checks}
\noindent A human QR does not rely solely on \(t\)-statistics; they evaluate alpha quality through multiple lenses \citep{de2018advances}. This skepticism is rooted in the "False Strategy Theorem," which warns that most backtested discoveries are merely artifacts of multiple testing on finite datasets. Our agent emulates this skepticism by incorporating a multi-objective promotion gate, shifting the focus from mere curve-fitting to the identification of robust statistical properties:

\begin{itemize}
    \item \textbf{Information Redundancy:} Using a memory-update mechanism, the agent ensures that new factors provide significant marginal information over existing benchmarks (e.g., the Fama-French five-factor model \cite{fama2015five}). This approach aligns with the principle of ``Feature Importance'' over "Backtest Performance" \cite{de2018advances}, ensuring the agent seeks unique structural drivers of returns. By filtering for marginal information gain, the framework avoids the ``substitution effect'' where redundant features inflate in-sample significance without adding genuine predictive value.
    
    \item \textbf{Implementation Feasibility:} Factors are penalized for excessive turnover and rapid alpha decay. This ensures the discovered signals are captureable in practice and not just artifacts of high-frequency noise. This reflects the ``Economic Reality'' constraint emphasized by \cite{de2018advances}, which argues that a valid strategy must possess a clear economic rationale and survive the frictions of market microstructure, rather than exploiting non-tradable statistical anomalies.\footnote{According to \cite{de2018advances}, the high failure rate of quantitative strategies often stems from "p-hacking" during the discovery phase. Our multi-objective gate acts as a functional heuristic for the Deflated Sharpe Ratio, implicitly accounting for the hidden trials involved in automated factor search by imposing stringent non-performance hurdles.}
\end{itemize}
\subsection{Strict Temporal and Policy Isolation}
\noindent To formally quantify the extent of data-mining, we adhere to a strict Out-of-Sample (OOS) protocol. All factor discovery and agent learning are finalized using data prior to December 2020. During the OOS period (2021--2024), the agent's internal state and factor library are frozen. This blind test ensures that the performance reported is not a result of look-ahead bias or iterative over-fitting to recent market regimes.

\subsection{The Anatomy and Economic Logic of Agent-Generated Factors}
\noindent This section provides an analysis of the factors synthesized by the autonomous agent, examining both their structural architecture and their underlying financial rationale. By transitioning from human-led prompting to an agentic discovery process, the framework identifies signals that are not only statistically robust but also deeply rooted in market microstructure and behavioral finance. We first delineate the self-iterative mechanism of autonomous synthesis and subsequently provide a thematic taxonomy of the discovered factors, demonstrating how the framework bridges the gap between machine-driven alpha extraction and economic interpretability.

\subsubsection{Mechanisms of Autonomous Factor Synthesis}
\noindent The evolution of factor discovery has moved progressively toward reducing human bias and increasing computational autonomy. This trajectory began with the application of machine learning to automate factor construction. For instance, \cite{fang2020neural} utilized deep learning architectures to capture complex non-linearities and temporal dependencies, moving beyond the limitations of traditional Genetic Programming (GP). More recently, the advent of Large Language Models (LLMs) introduced a semantic dimension to this automation. \cite{cheng2024gpt} and \cite{cheng2026large} demonstrated that LLMs could conceptualize innovative factors by leveraging their vast internal knowledge bases through human-led prompting, applying this logic across diverse datasets to extract alpha from textual and numerical insights.

Our methodology inherits the core strengths of these pioneering approaches—specifically the high-dimensional pattern recognition of neural frameworks and the creative hypothesis generation of LLMs. However, we propose a paradigmatic shift from these ``Traditional AI'' applications to an ``Agentic AI'' framework. This transition aligns with the emerging industry consensus, as noted by \cite{chen2025agentic}, who argues that the future of institutional asset management lies in the evolution from generative tools to autonomous agentic systems capable of independent data analysis and decision-making. 
While the neural approach of \cite{fang2020neural} remains tethered to fixed, human-engineered architectures, and the prompting methodology of \cite{cheng2024gpt} requires manual intervention to guide the model’s reasoning, our agentic framework operates with true functional autonomy. By moving beyond the "passive respondent" model, our approach realizes the "human-AI" collaborative paradigm envisioned in recent literature, where the AI agent functions as an autonomous researcher that systematically explores the factor zoo, rather than a mere tool for executing human-led prompts.

The defining advantage of this architecture is its role as a self-iterative researcher rather than a passive respondent. Unlike traditional models that function as opaque optimizers or "prompt-dependent" generators, our agent executes a continuous, closed-loop cycle of hypothesis formulation, empirical validation, and policy reflection. By utilizing a memory-update mechanism, the agent refines its search strategy based on accumulated backtesting evidence, allowing it to explore the "factor zoo" more systematically without being restricted by a specific network topology or a static set of instructions. This meta-autonomy ensures that the discovery process is not merely a product of pre-defined parameters but an evolving intelligence that adapts to market feedback, ensuring both statistical significance and economic persistence in the synthesized factors.

\subsubsection{Thematic Taxonomy and Financial Rationale: Economic Insights}
\noindent The factors synthesized by the autonomous agent exhibit a high degree of economic interpretability, focusing primarily on market microstructure, investor attention, and liquidity dynamics. As detailed in Exhibit \ref{tab:economic insights}, the agent demonstrates a sophisticated understanding of how turnover and flow shocks influence cross-sectional returns. A significant portion of the factor library is dedicated to capturing turnover-related anomalies across various temporal horizons. For instance, Factor 7 (Delayed Turnover Pressure) and Factor 10 (Medium Horizon Turnover Pressure) suggest that the agent has identified market crowding and capital rotation as primary drivers of price reversals or trend persistence.

The nature of these discovered signals offers a distinct advantage over the ``black-box'' outputs typical of the neural construction methods discussed by \cite{fang2020neural}. While neural-based frameworks are highly proficient at extracting alpha through high-dimensional feature mapping, the resulting signals are often embedded within opaque weight matrices, rendering them difficult for institutional investors to validate economically. In contrast, our agentic framework synthesizes transparent symbolic formulas that bridge the gap between high predictive power and rigorous interpretability. A unique analytical finding from the agent's output is the integration of execution frictions directly into the factor logic. Factor 2 (Friction Adjusted Flow Shock) exemplifies this by adjusting volume shocks for estimated trading costs, thereby prioritizing signals that remain profitable after accounting for market impact. This level of practical awareness is a direct result of the agent's iterative reflection on backtesting results, where it learns to penalize high-turnover signals that offer thin alpha cushions. Furthermore, factors such as Flow Volatility Imbalance (Factor 1) and Price Level Persistence (Factor 8) indicate that the agent can distinguish between steady buying interest and speculative noise, confirming that the autonomous discovery process aligns with established financial theories while identifying novel interaction effects.

\begin{table}[htbp]
\caption{\textbf{Autonomous Agentic AI-generated Factors and Economic Interpretation}}
\par
{\footnotesize
This table lists the constituent factors used in the composite factor construction and provides a concise economic interpretation for each factor.
The first column reports the factor index, the second column reports the factor name, and the third column summarizes the signal intuition and intended economic channel.
Factors are defined from price, trading activity, and turnover-related information and are used to predict next-day cross-sectional stock returns.
Descriptions are conceptual and implementation-neutral, intended to clarify variable meaning rather than to present performance evaluation.
}
\noindent
\begin{center}
\scriptsize
\begin{tabular*}{\textwidth}{@{\extracolsep{\fill}}rlp{0.55\textwidth}}
\toprule
No. & Factor name & Factor explanation \\
\midrule
3 & Composite Liquidity Demand & This factor combines turnover level and turnover acceleration to proxy for contemporaneous order-imbalance demand. Economically, stronger and broader buying pressure is more likely to reflect slow-moving information and limits-to-arbitrage, creating short-horizon return continuation. \\
8 & Defensive Mean-Reversion Signal & This factor is higher for stocks trading below recent trend anchors with relatively low realized risk. Economically, it targets underreacted, temporarily discounted names where mispricing correction can occur without extreme volatility risk. \\
12 & Delayed Flow Persistence & This factor focuses on lagged and smoothed turnover shocks, so it captures persistent order flow rather than one-day noise. Economically, it reflects gradual information diffusion and segmented liquidity provision. \\
11 & Flow Acceleration - Concave & This factor emphasizes persistent flow acceleration while down-weighting extreme spikes. Economically, it maps to investor attention waves where marginal price impact decays at very high trading activity levels. \\
1 & Flow Shock - Winsorized & This factor captures abnormal cross-sectional trading-demand shocks while limiting the influence of extreme outliers. Economically, it isolates broad attention or liquidity shocks that can move prices before full incorporation. \\
2 & Lagged Flow Pressure & This factor uses prior-day flow surprise to test whether liquidity-driven price pressure reverses or persists. Economically, persistence can arise when market makers and constrained arbitrageurs absorb inventory gradually. \\
9 & Medium-Horizon Attention & This factor measures sustained turnover over a medium horizon and proxies for persistent investor attention and crowding. Economically, prolonged attention can produce predictable demand pressure and delayed repricing. \\
5 & Persistent Turnover Intensity & This factor highlights relative liquidity-demand intensity across stocks over a short rolling horizon. Economically, names with persistently high turnover tend to carry stronger information arrival or speculative demand. \\
6 & Smoothed Flow Shock & This factor is a denoised flow-shock measure designed to suppress transitory spikes. Economically, it captures short-run demand pressure that survives smoothing filters and is therefore more likely to be priced with delay. \\
4 & Stable Turnover Trend & This factor rewards high but stable turnover and penalizes erratic liquidity bursts. Economically, stable participation often indicates institutional flow and lower adverse-selection uncertainty, supporting stronger signal reliability. \\
10 & Sustained Liquidity Attention & This factor captures persistent trading interest while preventing extremely active names from dominating the signal. Economically, it reflects gradual attention-driven capital reallocation rather than one-off noise trades. \\
7 & Turnover Volatility Risk & This factor measures short-window instability in liquidity demand and inventory risk transfer. Economically, high turnover volatility is linked to fragile liquidity states and time-varying risk premia. \\
\bottomrule
\end{tabular*}

\end{center}
\label{tab:economic insights}
\end{table}

\section{Robustness Tests}
\label{sec:robustness}
\noindent In this section, we conduct a comprehensive series of robustness checks to evaluate the practical viability and structural properties of our agentic framework. Specifically, our analysis spans four critical dimensions: (i) the persistence of predictive power over longer holding horizons, (ii) the economic viability of the strategy net of transaction costs, (iii) the stability of portfolio turnover, and (iv) the comparison between our autonomous agent and traditional AI configurations. Together, these tests show that the framework generates statistically significant and implementable alpha that survives real-world trading frictions, while cross-framework relative performance remains sensitive to factor-set construction and model specification.

\subsection{Longer-horizon predictability}
\noindent Exhibit \ref{tab:longer_horizon_predict} presents the out-of-sample performance of twelve individual Agentic-AI generated factors and their composite models over 1- to 7-day holding horizons. Both composite models deliver economically meaningful and statistically significant returns across horizons. At \(H1\), the Linear model reports an annualized return of 44.87\% (\(t=5.42\)), while the LGBM model reports 34.24\% (\(t=7.23\)). At \(H7\), the corresponding annualized returns are 18.95\% (\(t=6.20\)) for Linear and 12.90\% (\(t=7.61\)) for LGBM. These results indicate that machine-learning aggregation improves statistical stability, while return magnitude and inference strength can differ across model classes.

As expected, the predictive power generally decays as the holding horizon extends, with returns for most portfolios dropping monotonically from \(H1\) to \(H7\). However, the individual factors exhibit notable cross-sectional variation in their decay rates. While strong standalone predictors like Factors 6 and 10 show a steady decline in returns, their \(t\)-statistics remain highly significant across all seven days. Interestingly, Factors 5 and 9 exhibit lower initial returns but demonstrate remarkable persistence, with their statistical significance actually increasing over longer horizons (e.g., Factor 9's \(t\)-statistic rises from 3.80 at \(H1\) to 5.37 at \(H7\)), suggesting they capture slower-moving, longer-lasting market inefficiencies.

\begin{landscape}
\begin{table}[htbp]
\caption{\textbf{Long-Horizon Predictability of Factor Returns}}
\label{tab:longer_horizon_predict}
\par
{\footnotesize
This table reports the average daily (annualized) returns and associated $t$-statistics for hedge portfolios (Decile 10 minus Decile 1) formed based on the signals of twelve Agentic-AI generated factors and a composite linear model as outlined in Appendix \ref{app:aggregation_sorting}. At the end of each day, stocks are ranked into deciles based on their respective factor scores. We report the average daily performance for holding horizons ($H$) ranging from 1 day (1d) to 7 days (7d) after the portfolio formation. Returns are in percentages; \cite{newey1987hypothesis} adjusted $t$-statistics are in parentheses.

}
\begin{center}
\scriptsize
\begin{tabular*}{\linewidth}{@{\extracolsep{\fill}}lccccccc}
\toprule
Portfolio & H1 & H2 & H3 & H4 & H5 & H6 & H7 \\
\midrule
Linear   & 44.87 & 33.98 & 28.45 & 25.08 & 22.89 & 20.55 & 18.95 \\
         & (5.42) & (5.69) & (5.92) & (6.07) & (6.24) & (6.18) & (6.20) \\
LGBM     & 34.24 & 26.20 & 21.50 & 19.23 & 16.69 & 14.62 & 12.90 \\
         & (7.23) & (7.81) & (8.02) & (8.33) & (8.20) & (7.97) & (7.61) \\
Factor 1 & 16.33 & 7.91 & 5.35 & 3.73 & 3.70 & 2.51 & 1.83 \\
         & (5.71) & (4.06) & (3.56) & (2.94) & (3.30) & (2.49) & (1.95) \\
Factor 2 & -0.94 & -0.53 & -0.80 & 0.31 & -0.44 & -0.81 & -0.86 \\
         & (-0.35) & (-0.29) & (-0.55) & (0.25) & (-0.40) & (-0.80) & (-0.91) \\
Factor 3 & 21.95 & 12.93 & 9.64 & 8.22 & 7.24 & 5.83 & 4.82 \\
         & (4.82) & (4.16) & (3.91) & (4.01) & (4.05) & (3.68) & (3.36) \\
Factor 4 & 8.18 & 7.30 & 7.23 & 6.92 & 6.13 & 5.54 & 5.10 \\
         & (1.30) & (1.66) & (2.08) & (2.37) & (2.43) & (2.46) & (2.49) \\
Factor 5 & 11.64 & 8.42 & 6.97 & 5.99 & 5.55 & 5.75 & 5.72 \\
         & (2.81) & (2.98) & (3.10) & (3.10) & (3.29) & (3.85) & (4.24) \\
Factor 6 & 14.56 & 10.14 & 8.75 & 7.16 & 5.07 & 3.34 & 2.44 \\
         & (4.51) & (4.69) & (5.03) & (4.77) & (3.70) & (2.71) & (2.16) \\
Factor 7 & 5.42 & 3.97 & 3.82 & 2.91 & 2.34 & 1.97 & 2.53 \\
         & (1.43) & (1.53) & (1.91) & (1.72) & (1.57) & (1.43) & (1.99) \\
Factor 8 & 35.42 & 28.65 & 24.68 & 21.71 & 19.51 & 18.01 & 17.05 \\
         & (3.32) & (3.81) & (4.08) & (4.16) & (4.24) & (4.34) & (4.48) \\
Factor 9 & 12.59 & 10.00 & 8.83 & 7.68 & 6.88 & 6.56 & 6.63 \\
         & (3.88) & (4.42) & (4.88) & (4.99) & (5.05) & (5.25) & (5.75) \\
Factor 10& 10.51 & 8.93 & 8.20 & 7.50 & 6.88 & 6.31 & 6.00 \\
         & (2.88) & (3.50) & (4.04) & (4.35) & (4.53) & (4.64) & (4.88) \\
Factor 11& 8.87 & 5.58 & 3.84 & 3.07 & 3.36 & 3.11 & 3.32 \\
         & (1.24) & (1.10) & (0.95) & (0.89) & (1.11) & (1.15) & (1.35) \\
Factor 12& 5.49 & 2.39 & 0.90 & 0.47 & 0.28 & 0.09 & 0.56 \\
         & (1.67) & (0.99) & (0.47) & (0.29) & (0.20) & (0.07) & (0.49) \\
\bottomrule
\end{tabular*}
\end{center}
\end{table}
\end{landscape}

\subsection{Transaction Costs and Economic Viability}
\begin{figure}[htbp]
\begin{center}
\includegraphics[width=0.75\textwidth]{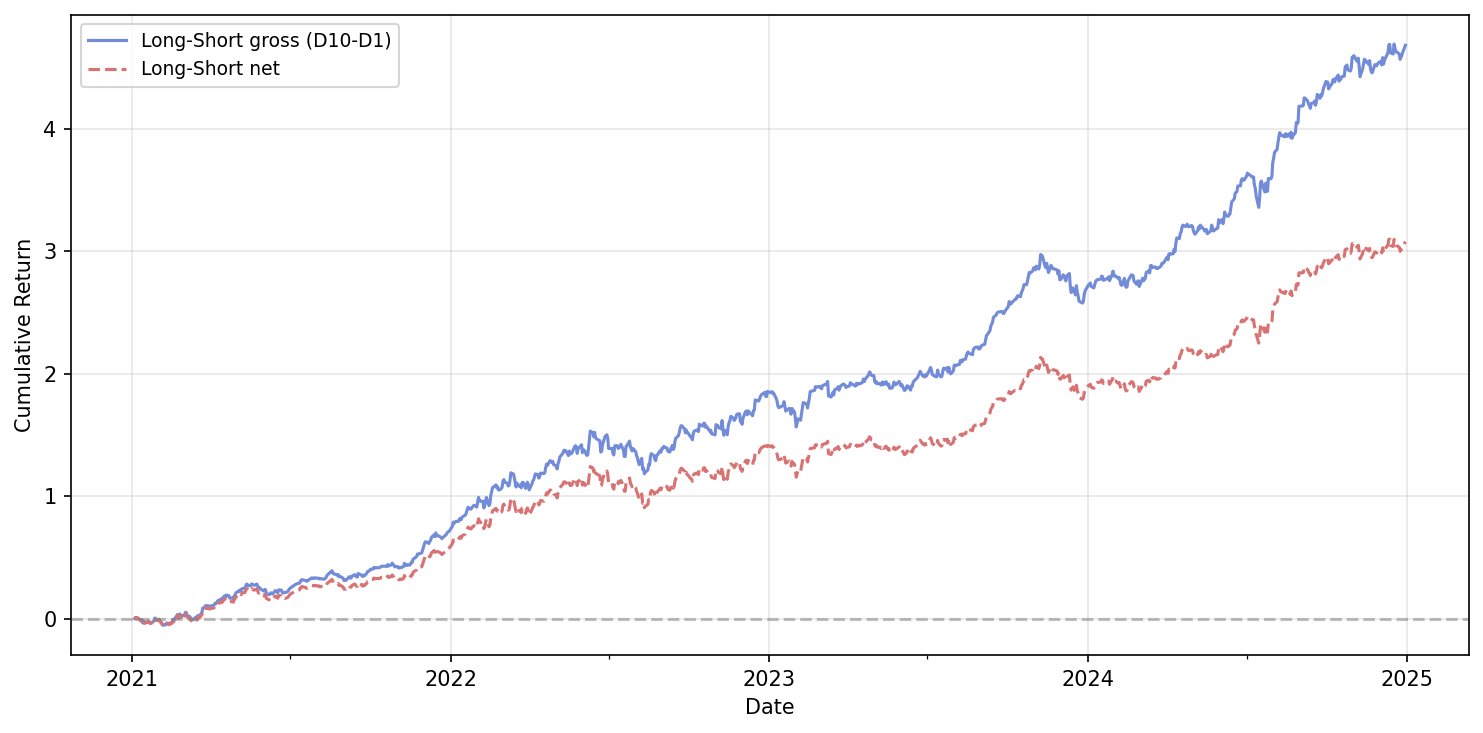}
\caption{\textbf{Out-of-Sample Cumulative Returns: Gross vs. Net Performance}}
\label{fig:multivariate_LS_cum}
\end{center}
{
\footnotesize
\noindent
\begin{spacing}{1.4}
This figure illustrates the out-of-sample cumulative performance of the composite long-short portfolio, constructed by aggregating multiple Agentic AI-generated factors as detailed in Appendix \ref{app:aggregation_sorting}. The blue solid line represents the gross cumulative return (top-minus-bottom deciles), while the red dashed line depicts the net cumulative return after accounting for transaction costs. The results are plotted over the main out-of-sample window from January 2021 to the end of the sample period.
\end{spacing}
}
\end{figure}

\noindent  A common criticism of high-frequency price-volume factors is that their alpha may be thin and easily eroded by trading frictions. We assume a linear transaction cost model of 3 basis points per dollar traded (one-way), which accounts for commission and spread. Exhibit \ref{fig:multivariate_LS_cum} presents the cumulative returns of the composite long-short strategy on both a gross and net basis from January 2021 to December 2024.
The results reveal a robust alpha cushion. While the gap between the blue (gross) and red (dashed, net) lines represents the cumulative impact of transaction costs, the net curve maintains a remarkably steady upward trajectory. Even after accounting for costs, the strategy achieves a substantial cumulative return (approximately 75\% net vs. 139\% gross). The fact that the net curve's slope remains consistently positive across different market regimes suggests that the agentic AI is not merely capturing transient noise, but is identifying structural premiums with sufficient magnitude to survive institutional-level execution costs.

\subsection{Turnover Analysis and Implementation Feasibility}
\noindent Exhibit \ref{tab:cost-turnover} details quarterly turnover and risk-adjusted performance. The strategy exhibits high daily turnover, ranging from 105.73\% to 114.43\%, reflecting the fast-decaying nature of the agent-discovered signals. Net returns remain positive in 14 of 16 quarters. In strong periods such as 2021Q4 and 2024Q2, the Net Sharpe ratio reaches 6.032 and 5.354, respectively. In the weakest quarter (2023Q1), the Net Sharpe is -0.075. Overall, the results indicate that the signal remains economically meaningful after implementation costs, while net performance is still time-varying across regimes.


\begin{table}[htbp]
\caption{\textbf{Quarterly Cost and Turnover Diagnostics (Out-of-Sample)}}\par
{\footnotesize
{} This table reports quarterly gross-versus-net performance together with turnover diagnostics for the composite long-short strategy. Avg Turnover denotes average daily long-short turnover within each quarter.
}
\noindent
\begin{center}
{\footnotesize
\begin{tabular*}{\linewidth}{@{\extracolsep{\fill}}lrrrrr}
\toprule
Quarter & Avg Turnover (\%) & Gross Ret (\%) & Net Ret (\%) & Gross Sharpe & Net Sharpe \\
\midrule
2021Q1 & 105.7314 & 9.3807 & 7.2873 & 1.7429 & 1.3914 \\
2021Q2 & 113.0857 & 12.6903 & 10.3115 & 3.2089 & 2.6491 \\
2021Q3 & 112.3283 & 12.1322 & 9.7434 & 3.6917 & 3.0085 \\
2021Q4 & 112.7979 & 23.7459 & 21.1027 & 6.7062 & 6.0322 \\
2022Q1 & 114.4251 & 21.5852 & 19.0321 & 3.7013 & 3.3114 \\
2022Q2 & 113.6247 & 18.3216 & 15.8532 & 3.1561 & 2.7742 \\
2022Q3 & 108.3577 & 1.6639 & -0.4296 & 0.3929 & 0.0426 \\
2022Q4 & 108.8101 & 13.5125 & 11.2048 & 2.6226 & 2.2123 \\
2023Q1 & 108.2341 & 1.2970 & -0.7221 & 0.3805 & -0.0752 \\
2023Q2 & 108.4066 & 3.0243 & 0.9678 & 1.1675 & 0.4135 \\
2023Q3 & 109.3437 & 18.0783 & 15.6682 & 6.0817 & 5.3309 \\
2023Q4 & 108.6696 & 4.5059 & 2.3825 & 1.2451 & 0.6996 \\
2024Q1 & 109.9185 & 6.2288 & 4.1142 & 2.2884 & 1.5450 \\
2024Q2 & 111.1760 & 17.2556 & 14.8230 & 6.1602 & 5.3539 \\
2024Q3 & 109.9314 & 17.6036 & 15.1524 & 4.2061 & 3.6689 \\
2024Q4 & 109.5543 & 5.4931 & 3.3327 & 1.9548 & 1.2195 \\
\bottomrule
\end{tabular*}

}
\end{center}
\label{tab:cost-turnover}
\end{table}

\subsection{Comparison of Agentic vs. Traditional AI Frameworks}
\noindent To isolate the source of alpha, we compare factors generated by the agentic framework against those from traditional AI factor mining. For each factor set, we apply two aggregation methods: a simple linear combination and a LightGBM (LGBM) integrator as detailed in Appendix \ref{app:aggregation_sorting}. 
Exhibit \ref{fig:multi_llm_portfolio_cumret} shows the out-of-sample cumulative long-short returns for these four combinations from 2021 to 2024.

\begin{figure}[htbp]
\begin{center}
\includegraphics[width=0.75\textwidth]{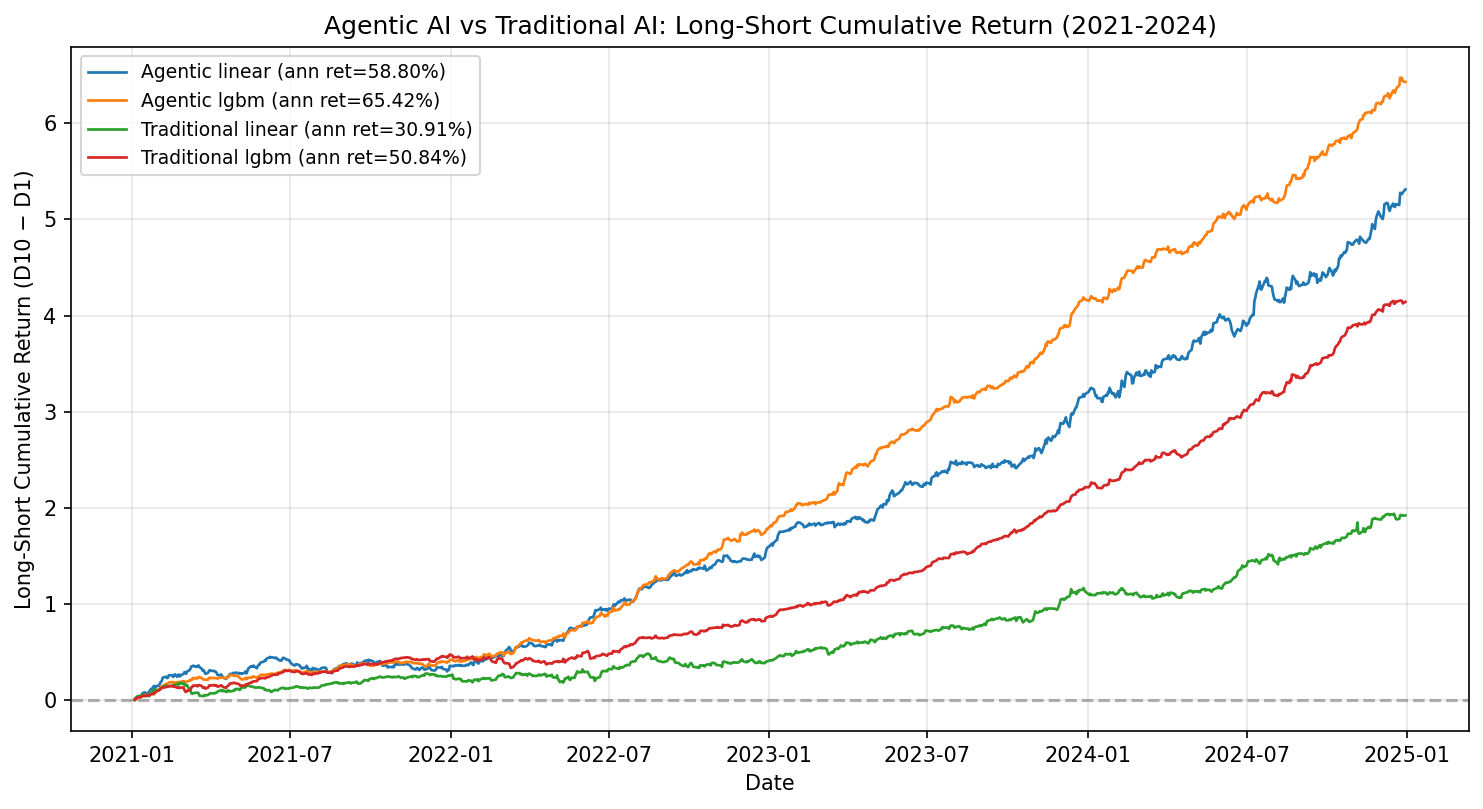}
\caption{\textbf{Out-of-Sample Comparison: Agentic vs. Traditional Factor Frameworks}}
\label{fig:multi_llm_portfolio_cumret}
\end{center}
{
\footnotesize
\noindent
\begin{spacing}{1.4}
This figure compares out-of-sample long--short cumulative returns (\(D10-D1\)) across four model specifications over 2021-01-01 to 2024-12-31 at daily frequency.
At each date, stocks are sorted by model-implied scores, with \(D10\) denoting the highest-score decile and \(D1\) the lowest-score decile; the long--short return is the daily return spread between the two deciles.
Each line reports the cumulative long--short return obtained by compounding daily spreads through time, and the legend reports the corresponding annualized return for each specification.
The horizontal axis is calendar time, the vertical axis is long--short cumulative return (\(D10-D1\)), and the dashed horizontal line marks the zero-return benchmark.
\end{spacing}
}
\end{figure} 

Exhibit~\ref{fig:multi_llm_portfolio_cumret} reports the out-of-sample cumulative long-short performance of four portfolio construction frameworks over the 2021--2024 period. Three main results emerge. First, the agentic framework outperforms the traditional benchmark under both aggregation schemes. Within the linear specification, the agentic portfolio achieves an annualized return of 58.80\%, compared with 30.91\% for the traditional linear portfolio. Within the nonlinear specification, the agentic LightGBM portfolio delivers the strongest overall performance, with an annualized return of 65.42\%, exceeding the 50.84\% achieved by the traditional LightGBM benchmark. This pattern indicates that the incremental value of the agentic pipeline is not confined to a specific downstream combiner, but instead reflects the superior quality of the underlying discovered signal set.

Second, the performance advantage of the agentic framework is persistent rather than episodic. The cumulative return curves begin to separate meaningfully from 2022 onward and remain well above their traditional counterparts through the end of the sample. This sustained spread suggests that the improvement is not driven by a single short-lived market episode, but instead reflects a more stable enhancement in cross-sectional signal extraction and factor organization.

Third, the nonlinear aggregation layer adds further value once higher-quality factors are available. Among all four specifications, agentic LightGBM produces the highest cumulative return trajectory, followed by agentic linear, traditional LightGBM, and traditional linear. This ranking suggests a complementary relationship between factor discovery and model flexibility: better upstream factor generation improves the opportunity set, while nonlinear aggregation helps capture residual interaction effects across discovered signals. Overall, Exhibit 19 provides direct out-of-sample evidence that the agentic factor framework delivers a materially stronger and more scalable return profile than traditional factor construction pipelines.

\section{Conclusion}
\label{sec:Conclusion}
\noindent 
This paper presents an autonomous framework for systematic factor investing driven by agentic artificial intelligence. We shift the focus from static machine learning applications to a self-iterative research process. The system formulates factor hypotheses, evaluates them under a strict empirical protocol, and updates its search policy based on accumulated evidence. By requiring economic interpretability and enforcing strict temporal isolation, the framework systematically mitigates data mining risks and the proliferation of spurious signals typically associated with automated discovery.

Empirical evaluations in the US equity market confirm the validity of this methodology. The autonomously generated factors deliver statistically significant risk-adjusted returns that are not spanned by standard asset pricing models. When integrated through a linear aggregation model, the composite portfolio achieves an annualized Sharpe ratio of 2.75 and an annualized gross return of 54.81 percent in the main out-of-sample window. These results remain robust after accounting for market microstructure frictions, transaction costs, and turnover constraints. Comparative tests indicate that cross-framework performance is sensitive to factor set composition and model choice across market regimes.

The evidence indicates that autonomous agentic systems offer a practical and scalable solution for modern quantitative asset management. By combining computational efficiency with scientific discipline, the framework provides a transparent method for discovering persistent market anomalies. Future research can build upon this foundation by expanding the empirical universe to include fundamental accounting data, macroeconomic indicators, and alternative asset classes, broadening the scope of autonomous financial research.~\\

\newpage
\noindent \textbf{Acknowledgement}

\noindent The authors thank Ivan Blanco (CUNEF), Kaiqi Hu (Rutgers Business School),  Bolong Wang (CITIC Securities), Yifan Ye (BNBU), Chao Zhang (HKUST Guangzhou), Yi Zhang (HKUST Guangzhou), and Yibin Zhang (Bosera Asset Management), and internal seminar participants at X Asset Management for helpful discussions, comments, and support.
These discussions, particularly those drawing upon practical industry insights, were conducted solely for academic purposes; the views expressed are those of the individuals and do not necessarily represent their employers.
The authors also thank the editorial teams of QuantML and LLMQuant for their insightful coverage and summary of this research across their leading Chinese practitioner-oriented social media platforms, which helped facilitate broader academic and industry exchange.
Any remaining errors or oversights are the responsibility of the authors.
~\\

\noindent \textbf{Disclosure of interest}

\noindent There are no interests to declare.
~\\

\noindent \textbf{Data availability}

\noindent Data will be made available on request. Additional interactive results, methodological documentation, and replication details are available at the project homepage \url{https://allenh16.github.io/agentic-factor-investing/}.

\newpage
\nocite{}
\small
\bibliography{ref}

@article{ye2025modeling,
  title={Modeling the Implied Volatility Smirk in {China}: Do Non-Affine Two-Factor Stochastic Volatility Models Work?},
  author={Ye, Yifan and Fan, Zheqi and Ruan, Xinfeng},
  journal={Journal of Futures Markets},
  volume={45},
  number={6},
  pages={612--636},
  year={2025},
  publisher={Wiley Online Library}
}

@article{he2025chronologically,
  title={Chronologically consistent large language models},
  author={He, Songrun and Lv, Linying and Manela, Asaf and Wu, Jimmy},
  journal={arXiv preprint arXiv:2502.21206},
  year={2025}
}

@article{brogaard2023machine,
  title={Machine learning and the stock market},
  author={Brogaard, Jonathan and Zareei, Abalfazl},
  journal={Journal of Financial and Quantitative Analysis},
  volume={58},
  number={4},
  pages={1431--1472},
  year={2023},
  publisher={Cambridge University Press}
}

@article{neely1997technical,
  title={Is technical analysis in the foreign exchange market profitable? A genetic programming approach},
  author={Neely, Christopher and Weller, Paul and Dittmar, Rob},
  journal={Journal of Financial and Quantitative Analysis},
  volume={32},
  number={4},
  pages={405--426},
  year={1997},
  publisher={Cambridge University Press}
}

@article{ding2023technical,
  title={Technical analysis as a sentiment barometer and the cross-section of stock returns},
  author={Ding, Wenjie and Mazouz, Khelifa and Ap Gwilym, Owain and Wang, Qingwei},
  journal={Quantitative Finance},
  volume={23},
  number={11},
  pages={1617--1636},
  year={2023},
  publisher={Taylor \& Francis}
}

@article{xu2025how,
  title={{AI} Agent Misinformation when Assisting Financial Decision-Making: Early Evidence From Stock Recommendations},
  author={Xu, Yongxin and Xuan, Yuhao and Zheng, Gaoping},
  journal={Available at SSRN 5651130},
  year={2025}
}

@article{fan2026deep,
  title={Deep surrogate for non-affine stochastic volatility option valuation models},
  author={Fan, Zheqi and Ruan, Xinfeng and Ye, Yifan},
  journal={Available at SSRN 6489158},
  year={2026}
}

@article{fan2025options,
  title={On Options-Driven Realized Volatility Forecasting: Information Gains via Rough Volatility Model},
  author={Fan, Zheqi and Wang, Meng Melody and Ye, Yifan},
  journal={Available at SSRN 5974814},
  year={2025}
}

@article{kong2024large1,
  title={Large language models for financial and investment management: Applications and benchmarks},
  author={Kong, Yaxuan and Nie, Yuqi and Dong, Xiaowen and Mulvey, John M and Poor, H Vincent and Wen, Qingsong and Zohren, Stefan},
  journal={Journal of Portfolio Management},
  volume={51},
  number={2},
  year={2024},
  publisher={With Intelligence}
}

@article{kong2024large2,
  title={Large Language Models for Financial and Investment Management: Models, Opportunities, and Challenges.},
  author={Kong, Yaxuan and Nie, Yuqi and Dong, Xiaowen and Mulvey, John M and Poor, H Vincent and Wen, Qingsong and Zohren, Stefan},
  journal={Journal of Portfolio Management},
  volume={51},
  number={2},
  year={2024}
}

@article{chen2022expected,
  title={Expected returns and large language models},
  author={Chen, Yifei and Kelly, Bryan T and Xiu, Dacheng},
  journal={Available at SSRN 4416687},
  year={2022}
}

@article{newey1987hypothesis,
  title={A simple, positive semi-definite, heteroskedasticity and autocorrelationconsistent covariance matrix},
  author={Newey, Whitney K and West, Kenneth D},
  journal={Econometrica},
  volume={55},
  number={3},
  pages={703--708},
  year={1987}
}

@article{chin2025technical,
  title={Technical indicators and the cross-section of corporate bond returns in a machine learning era},
  author={Chin, Jern Tat and Guo, Xu and Lin, Hai and Mei, Yi},
  journal={Journal of Financial Markets},
  pages={101029},
  year={2025},
  publisher={Elsevier}
}

@article{harvey2017presidential,
  title={Presidential address: The scientific outlook in financial economics},
  author={Harvey, Campbell R},
  journal={The Journal of Finance},
  volume={72},
  number={4},
  pages={1399--1440},
  year={2017},
  publisher={Wiley Online Library}
}

@article{chen2025chatgpt,
  title={{ChatGPT} and {DeepSeek}: Can they predict the stock market and macroeconomy?},
  author={Chen, Jian and Tang, Guohao and Zhou, Guofu and Zhu, Wu},
  journal={arXiv preprint arXiv:2502.10008},
  year={2025}
}

@article{chai2025generative,
  title={Generative {AI} for Finance: A New Framework},
  author={Chai, Bailin and Jiang, Fuwei and Meng, Lingchao and You, Tian and Zhou, Guofu},
  journal={Available at SSRN 6276278},
  year={2025}
}

@article{cheng2026large,
  title={Large Language Models and Futures Price Factors in {China}},
  author={Cheng, Yuhan and Liu, Yanchu and Zhou, Heyang},
  journal={Journal of Futures Markets},
  volume={46},
  number={2},
  pages={262--282},
  year={2026},
  publisher={Wiley Online Library}
}

@article{novy2016taxonomy,
  title={A taxonomy of anomalies and their trading costs},
  author={Novy-Marx, Robert and Velikov, Mihail},
  journal={The Review of Financial Studies},
  volume={29},
  number={1},
  pages={104--147},
  year={2016},
  publisher={Oxford University Press}
}

@article{tetlock2007giving,
  title={Giving content to investor sentiment: The role of media in the stock market},
  author={Tetlock, Paul C},
  journal={The Journal of Finance},
  volume={62},
  number={3},
  pages={1139--1168},
  year={2007},
  publisher={Wiley Online Library}
}

@article{lo2000foundations,
  title={Foundations of technical analysis: Computational algorithms, statistical inference, and empirical implementation},
  author={Lo, Andrew W and Mamaysky, Harry and Wang, Jiang},
  journal={The Journal of Finance},
  volume={55},
  number={4},
  pages={1705--1765},
  year={2000},
  publisher={Wiley Online Library}
}

@article{loughran2011liability,
  title={When is a liability not a liability? Textual analysis, dictionaries, and {10-Ks}},
  author={Loughran, Tim and McDonald, Bill},
  journal={The Journal of Finance},
  volume={66},
  number={1},
  pages={35--65},
  year={2011},
  publisher={Wiley Online Library}
}

@article{lopez2023can,
  title={Can {ChatGPT} forecast stock price movements? return predictability and large language models},
  author={Lopez-Lira, Alejandro and Tang, Yuehua},
  journal={arXiv preprint arXiv:2304.07619},
  year={2023}
}

@article{kelly2019characteristics,
  title={Characteristics are covariances: A unified model of risk and return},
  author={Kelly, Bryan T and Pruitt, Seth and Su, Yinan},
  journal={Journal of Financial Economics},
  volume={134},
  number={3},
  pages={501--524},
  year={2019},
  publisher={Elsevier}
}

@article{jegadeesh1993returns,
  title={Returns to buying winners and selling losers: Implications for stock market efficiency},
  author={Jegadeesh, Narasimhan and Titman, Sheridan},
  journal={The Journal of Finance},
  volume={48},
  number={1},
  pages={65--91},
  year={1993},
  publisher={Wiley Online Library}
}

@article{carhart1997persistence,
  title={On persistence in mutual fund performance},
  author={Carhart, Mark M},
  journal={The Journal of Finance},
  volume={52},
  number={1},
  pages={57--82},
  year={1997},
  publisher={Wiley Online Library}
}

@article{gu2020empirical,
  title={Empirical asset pricing via machine learning},
  author={Gu, Shihao and Kelly, Bryan and Xiu, Dacheng},
  journal={The Review of Financial Studies},
  volume={33},
  number={5},
  pages={2223--2273},
  year={2020},
  publisher={Oxford University Press}
}

@article{fama2015five,
  title={A five-factor asset pricing model},
  author={Fama, Eugene F and French, Kenneth R},
  journal={Journal of Financial Economics},
  volume={116},
  number={1},
  pages={1--22},
  year={2015},
  publisher={Elsevier}
}

@article{ko2025short,
  title={Short-Term Moving Average Distance and the Cross-Section of Stock Returns},
  author={Ko, Kuan-Cheng and Wang, Yanzhi and Yang, Nien-Tzu},
  journal={Financial Analysts Journal},
  volume={81},
  number={4},
  pages={121--141},
  year={2025},
  publisher={Taylor \& Francis}
}

@article{harvey2020false,
  title={False (and missed) discoveries in financial economics},
  author={Harvey, Campbell R and Liu, Yan},
  journal={The Journal of Finance},
  volume={75},
  number={5},
  pages={2503--2553},
  year={2020},
  publisher={Wiley Online Library}
}

@article{harvey2021lucky,
  title={Lucky factors},
  author={Harvey, Campbell R and Liu, Yan},
  journal={Journal of Financial Economics},
  volume={141},
  number={2},
  pages={413--435},
  year={2021},
  publisher={Elsevier}
}

@book{de2018advances,
  title={Advances in financial machine learning},
  author={L{\'o}pez de Prado, Marcos},
  year={2018},
  publisher={John Wiley \& Sons}
}

@article{fabozzi2025implementing,
  title={Implementing {AI} Foundation Models in Asset Management: A Practical Guide.},
  author={Fabozzi, Francesco A and L{\'o}pez de Prado, Marcos},
  journal={Journal of Portfolio Management},
  volume={52},
  number={2},
  year={2025}
}

@article{giamouridis2017systematic,
  title={Systematic investment strategies},
  author={Giamouridis, Daniel},
  journal={Financial Analysts Journal},
  volume={73},
  number={4},
  pages={10--14},
  year={2017},
  publisher={Taylor \& Francis}
}

@article{wei2022chain,
  title={Chain-of-thought prompting elicits reasoning in large language models},
  author={Wei, Jason and Wang, Xuezhi and Schuurmans, Dale and Bosma, Maarten and Xia, Fei and Chi, Ed and Le, Quoc V and Zhou, Denny and others},
  journal={Advances in Neural Information Processing Systems (NeurIPS)},
  volume={35},
  pages={24824--24837},
  year={2022}
}

@article{blitz2023beyond,
  title={Beyond {Fama-French} factors: Alpha from short-term signals},
  author={Blitz, David and Hanauer, Matthias X and Honarvar, Iman and Huisman, Rob and van Vliet, Pim},
  journal={Financial Analysts Journal},
  volume={79},
  number={4},
  pages={96--117},
  year={2023},
  publisher={Taylor \& Francis}
}

@article{yao2023react,
  title={{ReAct}: Synergizing Reasoning and Acting in Language Models},
  author={Yao, Shunyu and Zhao, Jeffrey and Yu, Dian and Du, Nan and Shafran, Izhak and Narasimhan, Karthik and Cao, Yuan},
  journal={International Conference on Learning Representations (ICLR)},
  year={2023},
  pages={1--33},
}

@article{chen2025cross,
  title={On Cross-Stock Predictability of Peer Return Gaps in {China}},
  author={Chen, Yilin and Fan, Zheqi},
  journal={Finance Research Open},
  volume={2},
  number={1},
  pages={100088},
  year={2026},
  publisher={Elsevier}
}

@article{rapach2019industry,
  title={Industry return predictability: A machine learning approach},
  author={Rapach, David E and Strauss, Jack K and Tu, Jun and Zhou, Guofu},
  journal={The Journal of Financial Data Science},
  volume={3},
  pages={9},
  year={2019}
}

@article{choi2025alpha,
  title={Alpha go everywhere: Machine learning and international stock returns},
  author={Choi, Darwin and Jiang, Wenxi and Zhang, Chao},
  journal={The Review of Asset Pricing Studies},
  volume={15},
  number={3-4},
  pages={288--331},
  year={2025},
  publisher={Oxford University Press}
}

@article{fama1993common,
  title={Common risk factors in the returns on stocks and bonds},
  author={Fama, Eugene F and French, Kenneth R},
  journal={Journal of Financial Economics},
  volume={33},
  number={1},
  pages={3--56},
  year={1993},
  publisher={Elsevier}
}

@article{brock1992simple,
  title={Simple technical trading rules and the stochastic properties of stock returns},
  author={Brock, William and Lakonishok, Josef and LeBaron, Blake},
  journal={The Journal of Finance},
  volume={47},
  number={5},
  pages={1731--1764},
  year={1992},
  publisher={Wiley Online Library}
}

@article{cheng2024gpt,
  title={{GPT}'s idea of stock factors},
  author={Cheng, Yuhan and Tang, Ke},
  journal={Quantitative Finance},
  volume={24},
  number={9},
  pages={1301--1326},
  year={2024},
  publisher={Taylor \& Francis}
}

@article{pu2026autonomous,
  title={Autonomous Market Intelligence: Agentic {AI} Nowcasting Predicts Stock Returns},
  author={Chen, Zefeng and Pu, Darcy},
  journal={Available at SSRN 6134446},
  year={2026}
}

@article{sharpe1964capital,
  title={Capital asset prices: A theory of market equilibrium under conditions of risk},
  author={Sharpe, William F},
  journal={The Journal of Finance},
  volume={19},
  number={3},
  pages={425--442},
  year={1964},
  publisher={Wiley Online Library}
}

@article{fang2020neural,
  title={Neural network-based automatic factor construction},
  author={Fang, Jie and Lin, Jianwu and Xia, Shutao and Xia, Zhikang and Hu, Shenglei and Liu, Xiang and Jiang, Yong},
  journal={Quantitative Finance},
  volume={20},
  number={12},
  pages={2101--2114},
  year={2020},
  publisher={Taylor \& Francis}
}

@article{cerniglia2020selecting,
  title={Selecting Computational Models for Asset Management: Financial Econometrics versus Machine Learning-Is There a Conflict?},
  author={Cerniglia, Joseph A and Fabozzi, Frank J},
  journal={The Journal of Portfolio Management},
  volume={47},
  number={1},
  pages={107--118},
  year={2020}
}

@article{chen2025agentic,
  title={Agentic {AI} and the Future of Institutional Asset Management.},
  author={Chen, Mike},
  journal={Journal of Portfolio Management},
  volume={51},
  number={10},
  year={2025}
}

@article{harvey2016cross,
  title={… and the cross-section of expected returns},
  author={Harvey, Campbell R and Liu, Yan and Zhu, Heqing},
  journal={The Review of Financial Studies},
  volume={29},
  number={1},
  pages={5--68},
  year={2016},
  publisher={Oxford University Press}
}

@article{han2016trend,
  title={A trend factor: Any economic gains from using information over investment horizons?},
  author={Han, Yufeng and Zhou, Guofu and Zhu, Yingzi},
  journal={Journal of Financial Economics},
  volume={122},
  number={2},
  pages={352--375},
  year={2016},
  publisher={Elsevier}
}

@article{han2013new,
  title={A new anomaly: The cross-sectional profitability of technical analysis},
  author={Han, Yufeng and Yang, Ke and Zhou, Guofu},
  journal={Journal of Financial and Quantitative Analysis},
  volume={48},
  number={5},
  pages={1433--1461},
  year={2013},
  publisher={Cambridge University Press}
}

@article{kelly2024virtue,
  title={The virtue of complexity in return prediction},
  author={Kelly, Bryan and Malamud, Semyon and Zhou, Kangying},
  journal={Journal of Finance},
  volume={79},
  number={1},
  pages={459--503},
  year={2024},
  publisher={Wiley Online Library}
}

@article{kelly2023financial,
  title={Financial machine learning},
  author={Kelly, Bryan and Xiu, Dacheng},
  journal={Foundations and Trends{\textregistered} in Finance},
  volume={13},
  number={3-4},
  pages={205--363},
  year={2023},
  publisher={Emerald Publishing Limited Boston—Delft}
}

@article{avramov2023machine,
  title={Machine learning vs. economic restrictions: Evidence from stock return predictability},
  author={Avramov, Doron and Cheng, Si and Metzker, Lior},
  journal={Management Science},
  volume={69},
  number={5},
  pages={2587--2619},
  year={2023},
  publisher={INFORMS}
}

@article{giglio2022factor,
  title={Factor models, machine learning, and asset pricing},
  author={Giglio, Stefano and Kelly, Bryan and Xiu, Dacheng},
  journal={Annual Review of Financial Economics},
  volume={14},
  number={1},
  pages={337--368},
  year={2022},
  publisher={Annual Reviews}
}

@article{cochrane2011presidential,
  title={Presidential address: Discount rates},
  author={Cochrane, John H},
  journal={The Journal of Finance},
  volume={66},
  number={4},
  pages={1047--1108},
  year={2011},
  publisher={Wiley Online Library}
}

\newpage
\begin{appendices}

\section{A Conceptual Framework for Agentic Factor Discovery}
\label{app:agentic_framework}
\noindent The methodology in this study extends traditional automated machine learning by adopting an autonomous agent-based framework. Unlike conventional methods that rely on brute-force combinatorial search and are consequently susceptible to \(p\)-hacking, this approach replicates the iterative heuristic and disciplined hypothesis testing of a quantitative researcher. The framework integrates logical reasoning \citep{wei2022chain} with sequential action execution \citep{yao2023react}. By requiring the formulation of an economic rationale prior to empirical validation, the system is structured to mitigate the risk of data-mining in high-dimensional factor spaces.

\subsection{The Reasoning-Action (ReAct) Framework}
The core logic of the factor discovery agent follows the ReAct paradigm \citep{yao2023react}, which interleaves reasoning traces and task-specific actions. In our framework, the agent's decision-making process at each step \(t\) is formalized as:
\begin{equation}
\label{eq:react_process}
(r_t, a_t) \sim \pi(s_t, h_t)
\end{equation}
where \(r_t\) is a reasoning trace (the "thought" or economic rationale), \(a_t\) is the action (the generated factor code), \(s_t\) is the current state of the factor library, and \(h_t\) is the historical trajectory of previous attempts. 

By incorporating ``Chain-of-Thought'' (CoT) prompting \citep{wei2022chain}, the agent is required to verbalize an economic rationale \(r_t\) before producing the mathematical expression \(a_t\). This ensures that the search space is constrained to factors that are \textit{a priori} economically plausible. The probability of generating a specific action can be conceptually decomposed as shown in Eq. \ref{eq:cot_decomposition}:

\begin{equation}
\label{eq:cot_decomposition}
P(a_t | s_t) = \sum_{r_t} P(a_t | r_t, s_t) P(r_t | s_t)
\end{equation}
The decomposition in Eq. \ref{eq:cot_decomposition} illustrates that the probability of discovering a high-quality factor \(a_t\) is conditioned on the quality of the latent reasoning \(r_t\), thereby acting as a structural regularizer against data-mining and spurious correlations.

\subsection{Interaction with the Backtesting Environment}
Following the execution of action \(a_t\), the agent receives an observation \(o_t\) from the environment \(\mathcal{E}\) (e.g., Sharpe ratio, information coefficient, or execution errors). The state for the next iteration is updated via a transition function \(\mathcal{T}\) defined in Eq. \ref{eq:state_transition}:

\begin{equation}
\label{eq:state_transition}
s_{t+1} = \mathcal{T}(s_t, a_t, o_t)
\end{equation}

This closed-loop system allows the agent to adjust its subsequent reasoning \(r_{t+1}\) based on the empirical performance of \(a_t\). The objective of the agent is to maximize the expected risk-adjusted utility of the discovered factor set over a finite horizon \(T\), as formulated in Eq. \ref{eq:objective_function}:
\begin{equation}
\label{eq:objective_function}
\max_{\pi} \mathbb{E}_{\pi} \left[ \sum_{t=1}^{T} \gamma^t R(s_t, a_t, o_t) \right]
\end{equation}
where \(R(\cdot)\) is the reward function and \(\gamma\) is the discount factor. 

\subsection{Note on the Conceptual Framework}
It is important to emphasize that Eq. \ref{eq:react_process} through Eq. \ref{eq:objective_function} serve as a conceptual framework to formalize the generative process of the Large Language Model (LLM), rather than an analytically tractable econometric model. In our implementation, the probability distributions, such as \(P(r_t | s_t)\) in Eq. \ref{eq:cot_decomposition}, are implicitly parameterized by the pre-trained weights of the LLM and the contextual prompts, rather than being explicitly estimated via maximum likelihood methods. The value of this formalization lies in illustrating how the agentic workflow simulates the scientific method—formulating hypotheses, executing empirical tests, and updating beliefs—to mitigate the risk of \(p\)-hacking in high-dimensional factor spaces.

\section{Factor Aggregation and Portfolio Construction}
\label{app:aggregation_sorting}
\noindent The transition from individual factor discovery to portfolio construction follows a structured ``forecast-then-sort'' framework. The methodology for synthesizing high-dimensional signals, constructing rank-based portfolios, and evaluating their economic significance follows the established literature \citep{han2016trend, rapach2019industry,choi2025alpha}.\footnote{While these studies share the same ``forecast-then-sort'' philosophy, the underlying predictive engines vary: \citet{han2016trend} employ linear cross-sectional regressions for trend signals, \citet{rapach2019industry} apply machine learning to industry-level sorting, and \citet{choi2025alpha} utilize machine learning frameworks to evaluate the economic value of discovered alphas.} 
This section details the technical implementation of these steps.

\subsection{Signal Synthesis and Expected Return Proxies}
For each asset \(i\) at time \(t\), we aggregate the information from the \(M\) discovered factors \(\mathbf{x}_{i, t}\) into a single composite score, \(S_{i, t}\). This score serves as our proxy for the conditional expected return in the next period:
\begin{equation}
S_{i, t} \equiv \hat{E}[r_{i, t+1} | \mathbf{x}_{i, t}] = \mathcal{M}(\mathbf{x}_{i, t}; \Theta)
\end{equation}
where \(\mathcal{M}(\cdot)\) represents the predictive aggregation model. 
This formulation provides a flexible alternative to heuristic aggregation methods. For instance, \citet{blitz2023beyond} argues for the robustness of simple equal-weighted Z-score averaging,\footnote{Specifically, the equal-weighted composite score is defined as: 
\begin{equation} 
S_{i, t} = \frac{1}{M} \sum_{m=1}^{M} z_{i, t, m} 
\end{equation} 
where \(z_{i, t, m}\) is the cross-sectionally standardized value of factor \(m\). 
This approach implicitly assumes uniform predictive power across all signals and ignores their covariance structure to minimize estimation risk.} noting that it effectively mitigates overfitting and selection bias in high-turnover signal contexts. While such a parsimonious approach avoids the estimation risks inherent in complex models, our predictive framework \(\mathcal{M}(\cdot)\) is designed to account for the varying predictive strength and cross-correlation structure of the underlying factors, thereby mapping the high-dimensional signal space more directly to the economic scale of expected returns.
This approach effectively captures information across multiple investment horizons by condensing a high-dimensional set of signals into a singular predictive proxy.

\subsection{Cross-Sectional Sorting and Portfolio Formation}
To translate these continuous forecasts into a tradable strategy, we employ a non-parametric sorting procedure. In each period \(t\), we rank all assets in the cross-section based on their predicted returns \(S_{i, t}\) and partition the universe into \(P\) quantile portfolios:
\begin{equation}
\mathcal{P}_{p, t} = \{ i : \text{Rank}(S_{i, t}) \in \text{Quantile}_p \}, \quad p=1, \dots, P
\end{equation}
The portfolio return for each quantile is calculated as the weighted average of its constituents, \(R_{p, t+1} = \sum_{i \in \mathcal{P}_{p, t}} w_{i, t} r_{i, t+1}\). This procedure allows us to evaluate the monotonicity of returns across different levels of forecasted alpha.

\subsection{The Long-Short Spread and Economic Value}
The primary object of interest is the zero-investment long-short portfolio (High-minus-Low), which captures the return spread between the top and bottom quantiles:
\begin{equation}
R_{HML, t+1} = R_{P, t+1} - R_{1, t+1}
\end{equation}
The performance of such ranked portfolios provides a robust measure of the economic value of return predictability. We further evaluate this spread using risk-adjusted alphas to ensure that the agent's discovered signals extract information that is not spanned by existing common risk factors.

\section{Non-linear Factor Aggregation via LightGBM}
\label{app:lightgbm_detail}
To aggregate the high-dimensional factor candidates discovered by our agent, we employ the Light Gradient Boosting Machine (LightGBM) framework. As noted by recent emerging machine learning in asset pricing literature \citep{gu2020empirical,kelly2024virtue,choi2025alpha}, the cross-section of stock returns is characterized by complex non-linearities and high-dimensional interactions that traditional linear models are ill-equipped to capture. LightGBM addresses these challenges through a regularized, additive tree-based approach.

\subsection{The Boosting Estimator}
We model the conditional expected return $E[r_{i, t+1} | \mathbf{x}_{i, t}]$ as an ensemble of $K$ regression trees:
\begin{equation}
\hat{r}_{i, t+1} = \sum_{k=1}^{K} \eta f_k(\mathbf{x}_{i, t}; \theta_k)
\end{equation}
where $\eta \in (0, 1]$ is the learning rate (or "shrinkage" parameter) that controls the contribution of each individual tree, and $f_k(\cdot)$ is a decision tree with parameters $\theta_k$ (representing split variables, split points, and leaf weights). This additive structure allows the model to learn the predictive function sequentially, with each subsequent tree fitting the residuals of the previous ensemble.

\subsection{Regularized Objective and Splitting Logic}
The model is optimized by minimizing a loss function that balances fit and parsimony. At each iteration $k$, the new tree $f_k$ is chosen to minimize:
\begin{equation}
\mathcal{L}^{(k)} = \sum_{i,t} L(r_{i, t+1}, \hat{r}_{i, t+1}^{(k-1)} + f_k(\mathbf{x}_{i, t})) + \gamma J + \frac{1}{2}\lambda \sum_{j=1}^J w_j^2
\end{equation}
where $J$ is the number of leaves and $w_j$ are the leaf weights. The inclusion of $L_1$ and $L_2$ penalties ($\gamma$ and $\lambda$) is crucial in a finance context to prevent the model from over-fitting to idiosyncratic noise.

The optimal split for any node is determined by maximizing the gain in the second-order Taylor approximation of the loss function. For a potential split into left ($I_L$) and right ($I_R$) child nodes, the gain $G$ is defined as:
\begin{equation}
G = \frac{1}{2} \left[ \frac{(\sum_{i \in I_L} g_i)^2}{\sum_{i \in I_L} h_i + \lambda} + \frac{(\sum_{i \in I_R} g_i)^2}{\sum_{i \in I_R} h_i + \lambda} - \frac{(\sum_{i \in I} g_i)^2}{\sum_{i \in I} h_i + \lambda} \right] - \gamma
\end{equation}
where $g_i$ and $h_i$ are the first and second-order gradients of the loss function. This criterion ensures that the model prioritizes splits that offer the most significant reduction in forecast error relative to the increased model complexity.

\subsection{Handling High-Dimensional Factor Spaces}
LightGBM introduces two specific algorithmic innovations that are particularly beneficial for the "factor zoo" problem:
\begin{itemize}
    \item \textbf{Gradient-based One-Side Sampling (GOSS):} By down-sampling observations with small gradients and focusing on those with large gradients, GOSS ensures that the estimator is driven by the most informative (and often hardest to predict) states of the economy.
    \item \textbf{Exclusive Feature Bundling (EFB):} Financial factors often exhibit high degrees of collinearity. EFB bundles mutually exclusive features to reduce the effective dimensionality of the search space, which enhances the robustness of the split selection process.
\end{itemize}

\subsection{Implementation and Out-of-Sample Guardrails}
To ensure the economic validity of our results, we implement several safeguards:
\begin{enumerate}
    \item \textbf{Temporal Validation:} We use a rolling-window training scheme (e.g., training on years $1$ to $t$ to predict $t+1$). This ensures that the model only uses information available at the time of the forecast, strictly avoiding look-ahead bias.
    \item \textbf{Hyper-parameter Tuning:} The complexity of the trees (e.g., maximum depth, number of leaves) and the shrinkage rate ($\eta$) are tuned via time-series cross-validation. This process ensures that the model's non-linear capacity is appropriately scaled to the signal-to-noise ratio of the data.
    \item \textbf{Feature Importance:} Beyond the portfolio sorts, we analyze the "Gain" and "Frequency" of each factor within the LightGBM ensemble to identify which discovered signals are the primary drivers of the non-linear predictive power.
\end{enumerate}

\section{Addtional tables and figures}
\begin{figure}[htbp]
\begin{center}
\includegraphics[width=0.99\textwidth]{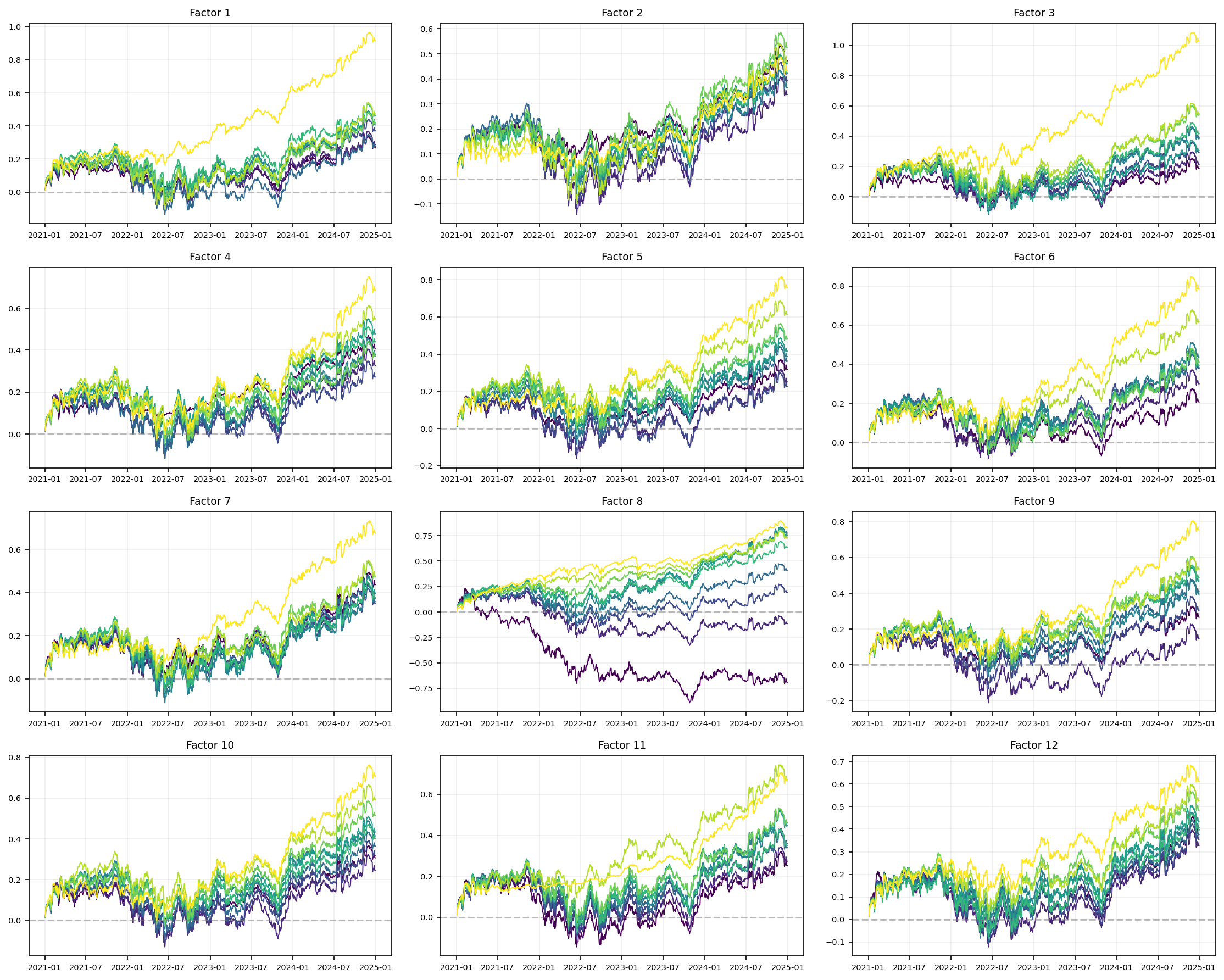}
\caption{\textbf{Univariate portfolio sort: Cumulative performance of ten decile portfolios}}
\label{fig:univariate_portfolio}
\end{center}
{
\footnotesize
\noindent
\begin{spacing}{1.4}
This figure reports univariate portfolio-sort results for Factor 1--Factor 11 over the out-of-sample period 2021-01-01 to 2024-12-31 at daily frequency.
For each date and factor, stocks are sorted into ten portfolios by cross-sectional factor scores, where \(D1\) is the lowest-score decile and \(D10\) is the highest-score decile; each decile return is the cross-sectional equal-weight mean of next-day returns \(y\).
Within each factor panel, the ten lines (\(D1\)--\(D10\)) plot cumulative decile performance transformed as \(\log(1+\mathrm{CumRet})\), where \(\mathrm{CumRet}_t=\prod_{\tau\le t}(1+r_\tau)-1\).
The horizontal axis is calendar time, the vertical axis is \(\log(1+\mathrm{CumRet})\), and the dashed horizontal line indicates the zero-cumulative-return benchmark.
\end{spacing}
}
\end{figure}

\begin{figure}[htbp]
\begin{center}
\includegraphics[width=0.99\textwidth]{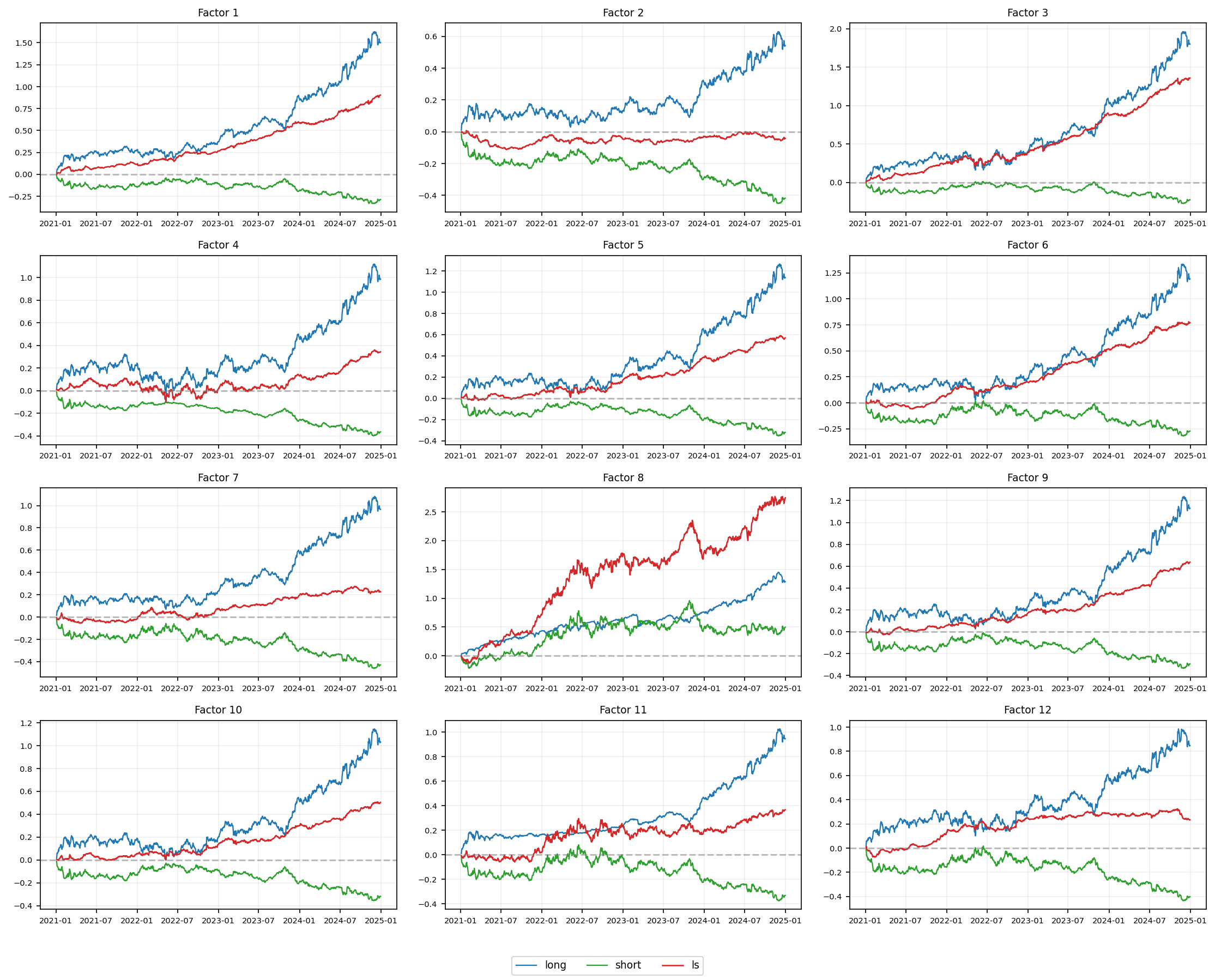}
\caption{\textbf{Univariate portfolio sort: Cumulative performance of long-short, long-only, short-only portfolios}}
\label{fig:univariate_portfolio_longshort}
\end{center}
{
\footnotesize
\noindent
\begin{spacing}{1.4}
This figure reports per-factor cumulative portfolio returns for Factor 1--Factor 11 over the test period 2021-01-01 to 2024-12-31 at daily frequency.
On each date, stocks are sorted into deciles by factor score in the cross-section; \(D10\) denotes the top-score decile and \(D1\) denotes the bottom-score decile, with decile daily returns defined as cross-sectional equal-weight means of next-day returns \(y\).
Within each factor panel, \textit{long} is the cumulative return of \(D10\), \textit{short} is the cumulative return of the short leg \((-D1)\), and \textit{ls} is the cumulative return of \(D10-D1\), each formed by compounding daily returns as \(\prod_t (1+r_t)-1\).
The horizontal axis is calendar time, the vertical axis is cumulative return, and the dashed horizontal line indicates the zero-return benchmark in every subplot.
\end{spacing}
}
\end{figure}

\begin{figure}[htbp]
\begin{center}
\includegraphics[width=0.99\textwidth]{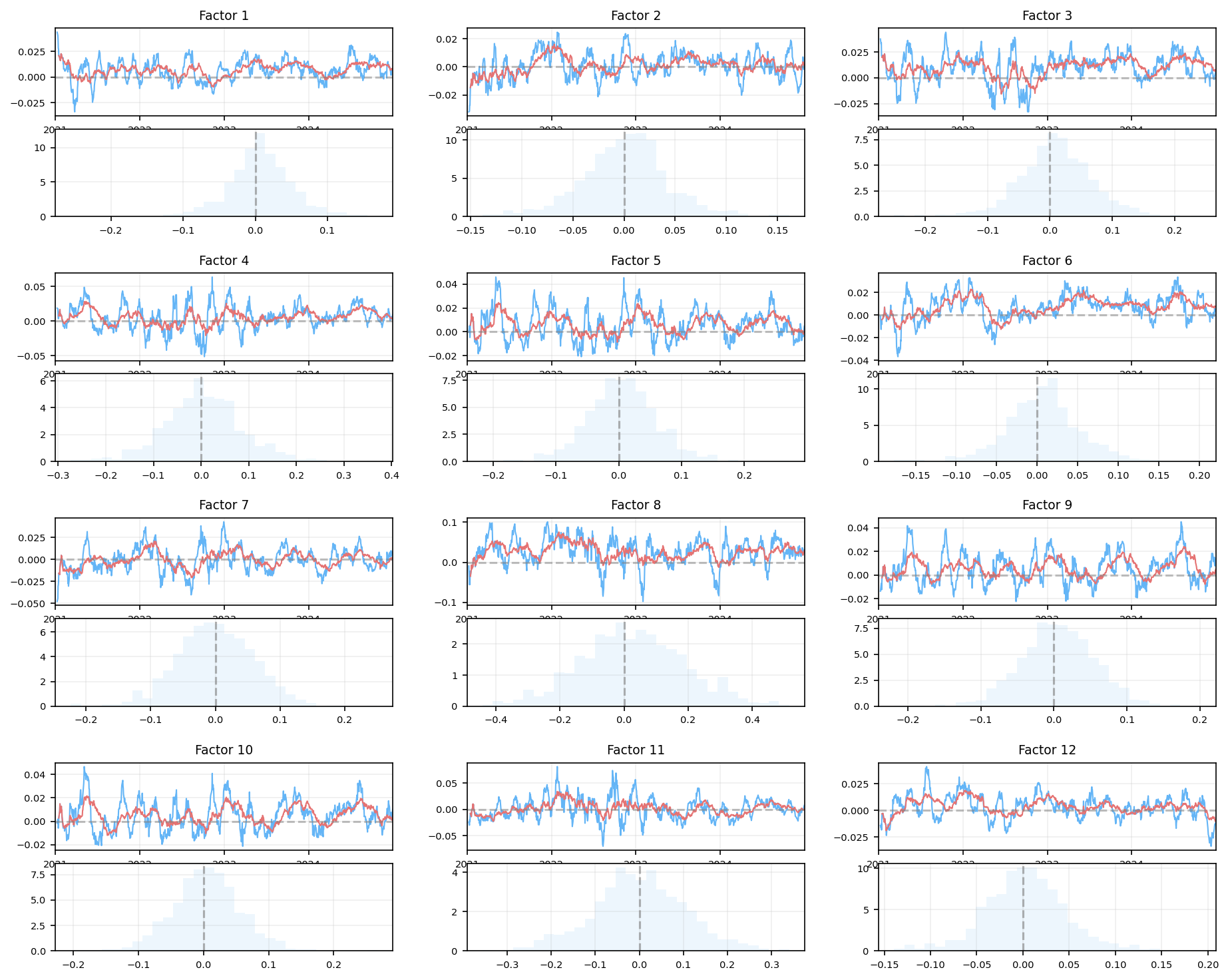}
\caption{\textbf{IC Diagnostics Grid: Top Rolling IC, Bottom Daily IC}}
\label{fig:univariate_ic_distribution}
\end{center}
{
\scriptsize
\noindent
\begin{spacing}{1.4}
This grid reports the Daily Rank IC for each factor. 
Upper subplots show 20-day and 60-day rolling means; lower subplots show the corresponding density distributions. 
The zero-benchmark is indicated by vertical dashed lines.
\end{spacing}
}
\end{figure}

\noindent This section provides supplementary figures to support the main empirical analysis. 
Exhibit \ref{fig:univariate_portfolio} illustrates the cumulative performance across ten decile portfolios for each factor generated by the autonomous system. 
Exhibit \ref{fig:univariate_portfolio_longshort} details the cumulative returns of the corresponding long-short, long-only, and short-only portfolio legs. 
Finally, Exhibit \ref{fig:univariate_ic_distribution} displays the diagnostic grid for the Information Coefficient (IC), featuring both rolling averages and daily distributions.

\end{appendices}


\end{document}